\documentclass[manuscript,acmsmall,screen]{acmart}

\renewcommand\footnotetextcopyrightpermission[1]{}
\AtBeginDocument{%
  \providecommand\BibTeX{{%
    \normalfont B\kern-0.5em{\scshape i\kern-0.25em b}\kern-0.8em\TeX}}}

\usepackage{multirow,makecell}
\usepackage{tcolorbox}
\usepackage{color,xcolor}
\usepackage{listings,amsfonts}
\usepackage{caption}
\usepackage{subcaption}
\usepackage{threeparttable}
\usepackage{bbding}
\usepackage{graphicx}
\usepackage{booktabs} 
\usepackage{longtable}
\usepackage{array}
\usepackage{url}

\AtEndPreamble{
	\usepackage{hyperref}
	\hypersetup{
		colorlinks = true,
		linkcolor = purple,
		anchorcolor = purple,
		citecolor = purple,
		filecolor = purple,
		urlcolor = purple
	}
}

\usepackage{tcolorbox}
\def\BibTeX{{\rm B\kern-.05em{\sc i\kern-.025em b}\kern-.08em
    T\kern-.1667em\lower.7ex\hbox{E}\kern-.125emX}}

\newcommand{\ea}{et~al.}

\begin{document}

\title{Software Engineering for Large Language Models: Research Status, Challenges and the Road Ahead}

\author{Hongzhou Rao}
\email{rhz@hust.edu.cn}
\authornotemark[1]
\affiliation{%
  \institution{Huazhong University of Science and Technology}
  \city{Wuhan}           
  \country{China}
}

\author{Yanjie Zhao}
\email{yanjie_zhao@hust.edu.cn}
\authornote{Co-first authors who contributed equally to this work.}
\affiliation{%
  \institution{Huazhong University of Science and Technology}
  \city{Wuhan}           
  \country{China}
}

\author{Xinyi Hou}
\email{xinyihou@hust.edu.cn}
\affiliation{%
  \institution{Huazhong University of Science and Technology}
  \city{Wuhan}           
  \country{China}
}

\author{Shenao Wang}
\email{shenaowang@hust.edu.cn}
\affiliation{%
  \institution{Huazhong University of Science and Technology}
  \city{Wuhan}           
  \country{China}
}

\author{Haoyu Wang}
\email{haoyuwang@hust.edu.cn}
\authornote{Haoyu Wang is the corresponding author (haoyuwang@hust.edu.cn).}
\affiliation{%
  \institution{Huazhong University of Science and Technology}
  \city{Wuhan}
  \country{China}
}

\begin{abstract}

The rapid advancement of large language models (LLMs) has redefined artificial intelligence (AI), pushing the boundaries of AI research and enabling unbounded possibilities for both academia and the industry.
However, LLM development faces increasingly complex challenges throughout its lifecycle, yet no existing research systematically explores these challenges and solutions from the perspective of software engineering (SE) approaches. To fill the gap, we systematically analyze research status throughout the LLM development lifecycle, divided into six phases: requirements engineering, dataset construction, model development and enhancement, testing and evaluation, deployment and operations, and maintenance and evolution. We then conclude by identifying the key challenges for each phase and presenting potential research directions to address these challenges. In general, we provide valuable insights from an SE perspective to facilitate future advances in LLM development.

\end{abstract}

\maketitle

\section{Introduction}

In recent years, large language models (LLMs) have advanced rapidly, leading to their performance exceeding human capabilities in certain domains~\cite{schoenegger2024wisdom,si2024can,zheng2024judging,gao2024llm,street2024llms}. Alongside this progress, emerging technologies such as AI-driven code generation~\cite{guo2024deepseek,zhu2024deepseek,lozhkov2024starcoder,hui2024qwen2}, multimodal models~\cite{sun2024generative,rasheed2024glamm}, and AI agents~\cite{xi2025rise} are also evolving at an unprecedented pace. These developments are rapidly expanding the role of LLMs in critical domains such as software development, healthcare~\cite{goyal2024healai,yang2024talk2care,gebreab2024llm}, and finance~\cite{yu2025fincon,zhao2024revolutionizing}. As LLMs become foundational infrastructure for general-purpose intelligence~\cite{mumuni2025large}, ensuring their reliability, efficiency, and adaptability is important as well as challenging for the software engineering (SE) community. Therefore, a systematic investigation into the development and engineering of LLMs is essential.

\textbf{The development of LLMs is a multifaceted process, from dataset preparation and model training to deployment and maintenance.} SE plays a central role throughout this lifecycle, offering foundational principles and methodologies to manage complexity, ensure robustness, and support scalability. These are increasingly embodied in an ecosystem of specialized tools and frameworks that facilitate each stage of the development process. For instance, Hugging Face provides tools such as Transformers~\cite{huggingface_transformers} for model training and the PEFT library~\cite{huggingface_peft} for efficient fine-tuning methods. Micros
ft’s DeepSpeed~\cite{deepspeed} enhances large-scale model training through deep learning optimizations, while OpenAI offers LLM APIs~\cite{openai_api_docs} that enable interaction with the GPT family models. Additionally, evaluation frameworks like LM-Eval-Harness~\cite{lm_evaluation_harness} and community-driven platforms such as the Open LLM Leaderboard~\cite{open_llm_leaderboard} offer standardized benchmarks. Development toolkits like LangChain~\cite{langchain}  modularize and streamline the construction of LLM-based applications. These tools embody SE principles and reflect the increasing need for structured, reliable, and scalable LLM development pipelines.

Beyond model training and development, SE also plays a critical role in optimizing the efficiency and scalability of LLM deployment. We can develop tools that integrate techniques such as model compression, quantization, and inference optimization to reduce latency and resource consumption. For example, TensorRT~\cite{nvidia_tensorrt} and vLLM~\cite{vllm} employ these techniques to enable cost-effective and efficient LLM inference in production environments. 
Furthermore, \textbf{specialized protocols like Model Context Protocol (MCP)~\cite{anthropic2025mcp} and Agent-to-Agent (A2A)~\cite{google2025a2a} standardize interactions between LLM-based agents, tools, and multi-agent systems, ensuring interoperability and streamlined integration in production pipelines}. SE practices play a crucial role in implementing these protocols, providing structured approaches to develop and maintain associated SDKs~\cite{openai_api_docs}, client libraries~\cite{huggingface_transformers, langchain}, and server components~\cite{tensorflow2025serving, azure_openai_service}, while ensuring code quality, reliability, and scalability.
In summary, from infrastructure development and data/model management to deployment and inference acceleration, SE methodologies and tools continue to shape the rapid evolution and widespread adoption of LLMs.

Despite these advancements, there are unique SE challenges for LLM development. High computational costs~\cite{stojkovic2024towards,shi2024efficient}, non-deterministic testing~\cite{song2024good}, and continuous model updates in dynamic environments~\cite{ma2024schrodinger} demand a re-evaluation of traditional SE practices. Traditional MLOps methodologies, initially designed for smaller-scale machine learning (ML) models, are no longer well-suited for LLMs, necessitating large language model operations (LLMOps)~\cite{diaz2024large}. Furthermore, LLMs deviate from traditional software paradigms: unlike traditional programs that produce deterministic outputs, LLMs generate responses probabilistically due to their neural network-based reasoning mechanisms. Additionally, their complex architectures and large scale lead to their outputs being challenging to explain, making interpretability and debugging far more difficult than in traditional software systems. These factors highlight the need for engineering solutions to address LLMs' inherent unpredictability, lack of transparency, and unique operational constraints.

However, we found that extensive research has explored LLM capabilities~\cite{hou2024large,wu2024survey,zhu2023large}, while systematic investigations from an SE perspective are still lacking. To address this gap, we present \textbf{the first comprehensive study of the SE challenges encountered throughout the LLM development lifecycle and outline future research directions}. Specifically, we categorize the LLM development lifecycle into six key phases: requirements engineering (RE), dataset construction, model development and enhancement, testing and evaluation, deployment and operations, and maintenance and evolution. 
For each phase, we analyze the current research status to identify key challenges and propose potential future research directions from an SE perspective.

In summary, our primary contributions are:
\begin{itemize}
    \item Our work is the first to investigate the role of SE in the development of LLMs, filling the gap in current research.
    \item We divide the LLM development lifecycle into six phases and systematically analyze the scope and significance of SE for LLMs.
    \item We analyze the latest research on LLMs, identify current challenges, and propose corresponding future research directions.
\end{itemize}

\begin{figure}[htbp!]  
    \centering     
    \includegraphics[width = 0.9\textwidth]{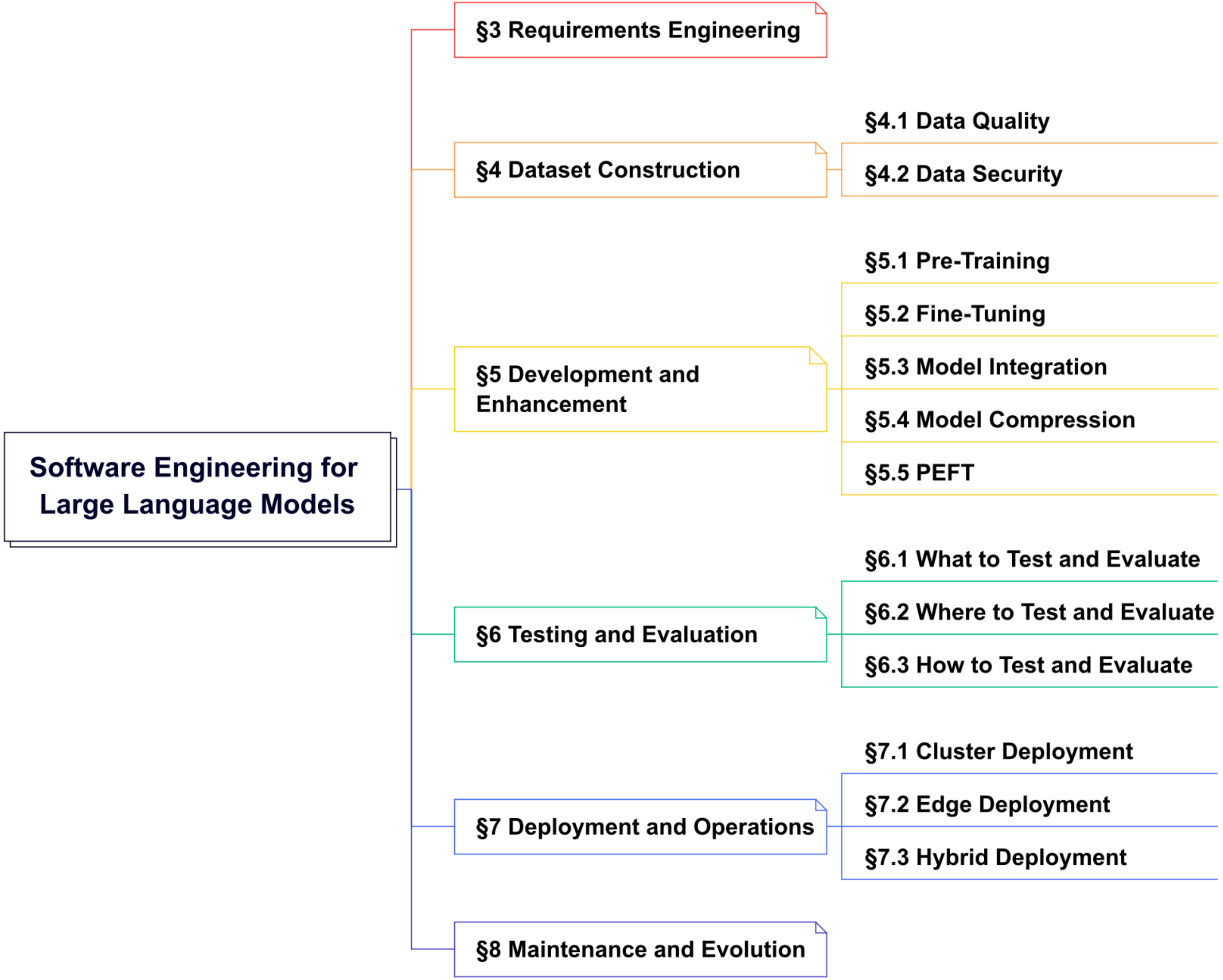}  
    \caption{Phase Organization Overview for the LLM Development Lifecycle.}
    \label{fig:se4llm_road_ahead}

\end{figure}

The remainder of this paper is structured as follows: In \S\ref{sec:def_sig}, we introduce the scope and significance of SE for LLMs. We then analyze various aspects of LLM development, covering RE (\S\ref{sec:re}), dataset construction (\S\ref{sec:datasets}), development and enhancement (\S\ref{sec:development_enhancement}), testing and evaluation (\S\ref{sec:testing_evaluation}), deployment and operations (\S\ref{sec:model_deployment}), and maintenance and evolution (\S\ref{sec:maintenance_evolution}), as illustrated in \autoref{fig:se4llm_road_ahead}. Finally, we introduce the related work in \S\ref{sec:related_work} and conclude the paper in \S\ref{sec:conclusion}.

\section{Scope and significance}\label{sec:def_sig}

In this section, we introduce the scope of SE for LLMs in \S\ref{sec:what_is_se4llm} and its significance in \S\ref{sec:why_is_se4llm}. 


\subsection{Scope}\label{sec:what_is_se4llm}

As summarized in \autoref{tab:software_vs_llm}, we can see that LLMs share similarities with traditional software while also exhibiting differences. Unlike traditional software, LLMs are built upon neural network architectures, resulting in non-deterministic outputs for identical inputs. Additionally, they differ from traditional software in terms of executability and testing methodologies. Despite these differences, LLMs retain many characteristics of traditional software.
Given these hybrid characteristics, SE methodologies can be applied to enhance LLM in phases such as development, deployment, and maintenance. To systematically investigate this intersection, we explore how SE methodologies support various phases of the LLM development lifecycle, which consists of the phases illustrated in \autoref{fig:llm_lifecycle}:

\begin{figure}[htbp]  
    \centering     
    \includegraphics[width = \textwidth]{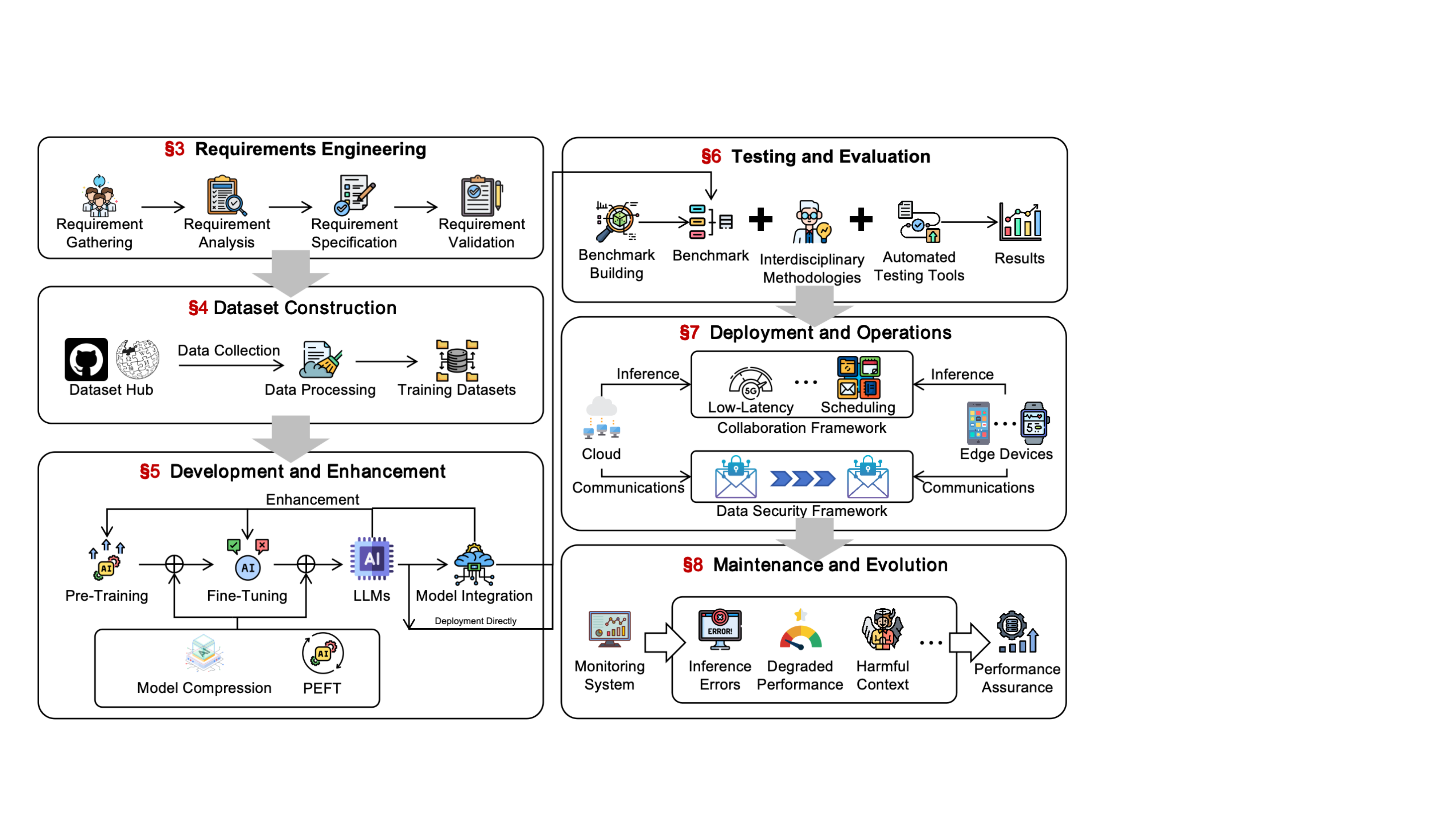}  
    \caption{Detailed Activity Breakdown for the LLM Development Lifecycle.}
    \label{fig:llm_lifecycle}
\end{figure}

\begin{itemize}
    \item \textbf{RE.} 
    The initial phase of RE for LLMs involves identifying specific performance metrics (e.g., accuracy, latency, energy consumption) and functional capabilities (e.g., reasoning, multimodal understanding) that the model is expected to possess, which is typically followed by a systematic process of requirement refinement, feasibility analysis, and validation to ensure that the specified goals are both realistic and implementable.
    
    \item \textbf{Dataset construction.} 
    Once requirements are established, vast datasets must be prepared for subsequent pre-training and fine-tuning. The construction of datasets involves data collection and processing to produce high-quality and secure datasets, as the dataset's quality significantly impacts model performance~\cite{liu2024datasets}. Harmful data can lead to unintended biases or malicious content generation.
    
    \item \textbf{Development and enhancement.} 
    The development of LLMs generally comprises two key stages: pre-training and fine-tuning. Building an LLM from scratch entails designing and implementing foundational architectures (e.g., Transformers) and performing large-scale training, which is a highly complex and resource-intensive engineering undertaking. Fortunately, open-source development frameworks facilitate this process. For example, Hugging Face Transformers~\cite{huggingface_transformers} facilitates model pre-training and fine-tuning, making it a mainstream tool for LLM development. After development, models can undergo further training to strengthen specific capabilities. Additionally, they can be integrated with tools to evolve into more advanced LLM agents.
    
    \item \textbf{Testing and evaluation.} 
    Evaluating an LLM requires comprehensive and systematic testing to assess its diverse capabilities across different tasks and scenarios. However, due to the inherent complexity of LLMs and their non-deterministic outputs, traditional software testing methodologies are often not suitable. When evaluating a model, it is essential to consider not only basic performance measures but also practical challenges, such as hallucinations, inconsistent outputs, context sensitivity, and other factors that may impact its reliability. Therefore, further research is needed to improve evaluation methods for LLMs.

    \item \textbf{Deployment and operations.} 
    Once validated, an LLM can be deployed across various application scenarios. Some models are hosted in cloud environments and accessed via APIs (e.g., GPT-4.5~\cite{openai2025gpt45}, Claude-3.7-Sonnet~\cite{anthropic2025claude37sonnet}), while others are deployed on edge devices or within hybrid edge–cloud setups to achieve low-latency and resource-efficient inference. However, this diversity in deployment environments introduces new challenges, ranging from scalability and reliability to resource allocation and system integration. Addressing these issues requires strong support from SE practices, such as automated deployment pipelines, environment-specific optimizations, and real-time monitoring systems.
    
    \item \textbf{Maintenance and evolution.} 
    During operation, LLMs require substantial computational resources and may encounter issues such as performance degradation, inference errors, or the need for retraining and knowledge updates. Therefore, LLMs also require systematic maintenance, bringing additional challenges beyond those of traditional software systems.
\end{itemize}

From this lifecycle perspective, it is evident that SE methodologies are deeply integrated into the construction, deployment, and utilization of LLMs. In dataset construction, specialized tools facilitate data cleaning and synthesis. Model development relies heavily on existing frameworks, while model enhancement often involves integrating LLMs with external tools to extend their functionality. Furthermore, testing, evaluation, and deployment require novel approaches distinct from traditional SE practices due to challenges such as debugging difficulties, output variability, and high computational demands. Beyond engineering complexities, LLMs introduce additional security and ethical concerns. They are vulnerable to adversarial threats, including data poisoning and prompt injection attacks, and may perpetuate biases inherent in their training data. Addressing these issues necessitates SE techniques. \textbf{Overall, SE in LLMs aims to support structured and efficient development across the full lifecycle—from requirements engineering and dataset construction to model deployment and maintenance—while also tackling concerns such as security, ethical responsibilities, and regulatory compliance.}

\begin{table}[]
\centering
\caption{Comparison of Characteristics Between Traditional Software and LLM.}
\label{tab:software_vs_llm}
\resizebox{\linewidth}{!}{%
\begin{tabular}{|l|p{0.45\linewidth}|p{0.45\linewidth}|c|}
\hline
\textbf{Characteristic} & \textbf{Traditional Software} & \textbf{LLM} & \textbf{Similarity} \\ \hline
Determinism & Producing consistent outputs for the same inputs. & Generating probabilistic outputs with inherent variation. & Low \\ \hline
Executability & Executing based on explicit, defined logic. & Processing through neural inference with limited transparency. & Medium \\ \hline
Maintainability & Maintaining through code modifications and debugging. & Improving via fine-tuning, retraining, or data augmentation. & High \\ \hline
Reusability & Reusing code components across different projects. & Adapting pre-trained models for various tasks. & High \\ \hline
Testability & Supporting systematic unit and integration testing. & Requiring output-based evaluation with uncertainty tolerance. & Medium \\ \hline
Scalability & Expanding through modular design principles. & Scaling via MoE, LoRA, RAG, and parameter-efficient methods. & High \\ \hline
Deployability & Requiring platform-specific deployment approaches. & Functioning across platforms with similar infrastructure needs. & High \\ \hline
\end{tabular}%
}
\end{table}

\subsection{Significance}\label{sec:why_is_se4llm}

LLMs are hard to understand because they have very complex structures and rely heavily on neural networks to make decisions. This lack of explanation makes it difficult to optimize, test, and maintain these models over time. They also introduce new security and privacy issues that are different from those in traditional software. As LLM applications expand across various domains and industries, these risks become increasingly critical. To address these issues, we require robust safety measures and transparent development processes.  Given these challenges, the application of SE methodologies can help ensure that LLMs are built, deployed, and maintained reliably and responsibly while addressing associated security, ethical, and regulatory concerns.

\subsubsection{SE for LLM Development}

SE provides systematic and automated tools and methodologies to support the development of LLM, significantly enhancing both efficiency and reliability. Due to the massive scale of LLMs and their long training cycles, manually managing parameters, datasets, and code versions is complex and error-prone. SE methodologies, such as LLMOps pipelines, automated hyperparameter management tools, and model version control systems (e.g., Weights \& Biases~\cite{wandb}), facilitate automation and standardization in the model development phase. These tools reduce the likelihood of human error and enhance reproducibility.

Beyond automation, LLM development presents challenges that differ from traditional SE, such as stability issues during pre-training~\cite{tang2023improving} and catastrophic forgetting during fine-tuning~\cite{li2024revisiting}. These issues can be mitigated through real-time training monitoring and content-detection techniques. Additionally, SE methodologies help manage the complexities introduced by parameter-efficient fine-tuning (PEFT). For example, when integrating adapter layers~\cite{kim2024hydra} or Low-Rank Adaptation (LoRA)~\cite {hayou2024lora+}, it is critical to ensure effective management, maintain compatibility between multiple tasks, and preserve performance stability. The principles of SE, such as modular design and continuous integration (e.g., automated adapter testing and parameter compatibility verification), provide structured solutions for efficiently managing and securely applying these fine-tuning techniques.

Furthermore, SE plays an essential role in model versioning and iterative upgrades. Automated tools that compare different model versions help prevent performance degradation and functionality loss, enabling smoother and more reliable updates. Thus, beyond enhancing development efficiency, SE methodologies also help address engineering challenges unique to LLMs, facilitating their continuous improvement and large-scale deployment.

\subsubsection{SE for LLM Deployment}

SE plays a crucial role in the deployment of LLMs by enabling efficient model compression~\cite{yao2025efficient} and automated deployment tools (e.g., Hugging Face Inference Endpoints~\cite{huggingface_inference_endpoints}). While LLMs offer superior performance, their substantial computational demands pose significant challenges for deployment in resource-constrained devices, such as mobile devices and IoT edge devices~\cite{dhar2024empirical}. SE addresses these challenges by advancing model compression techniques, including quantization~\cite{chee2024quip}, knowledge distillation (KD)~\cite{zhou2024teaching}, and pruning~\cite{zhao2024apt}, as well as adapter-based approaches such as LoRA~\cite{zhang2023adalora} and Prefix-Tuning~\cite{li2021prefix}, effectively reducing parameter size and computational cost. For instance, quantization techniques, such as INT8~\cite{dettmers2022gpt3} and INT4~\cite{lin2024qserve}, enable large models to achieve efficient inference on consumer-grade GPUs and even mobile devices~\cite{yin2024llm}, significantly expanding their applicability. Additionally, the development of modular and standardized LLM service interfaces (e.g., OpenAI API~\cite{openai_api}) allows developers to seamlessly deploy and transition between models across diverse environments, thereby reducing system deployment complexity.

Beyond model optimization, SE also enhances LLM deployment through the development of standardized interaction protocols, such as MCP~\cite{anthropic2025mcp} for structured tool integration and A2A~\cite{google2025a2a} communication for multi-agent collaboration. These protocols streamline integration with external APIs, databases, and distributed AI agents while ensuring interoperability and fault tolerance. Additionally, the development of modular and standardized LLM service interfaces (e.g., OpenAI API~\cite{openai_api}) allows developers to seamlessly deploy and transition between models across diverse environments, thereby reducing system deployment complexity. 

\subsubsection{SE for LLM Maintenance}

SE methodologies are essential for the long-term maintenance and evolution of LLMs. Once deployed, LLMs require continuous updates to incorporate new features, adapt to evolving business needs, and respond to environmental changes. As the number of model versions grows, managing version compatibility, tracking dataset evolution~\cite{ostendorff2024llm}, and ensuring API stability~\cite{ma2024my} become critical challenges. SE addresses these challenges through version control systems for models and datasets, modular architecture designs (e.g., LoRA adapters), and automated regression testing frameworks (e.g., LLM-specific continuous integration tools), enabling efficient tracking of model functionality changes and rapid issue resolution. For instance, when releasing new versions of the Gemini~\cite{google_gemini} and LLaMA~\cite{llama} families, Google and Meta employ rigorous SE practices to manage version compatibility, maintain API stability, and ensure seamless migration for downstream users. These engineering-driven maintenance strategies significantly enhance model evolution efficiency, ensuring consistent and reliable performance while supporting the sustained deployment of LLMs.

\subsubsection{SE for LLM Security}

LLMs process vast amounts of sensitive data during inference and deployment, particularly in API-based services, raising concerns about data leakage and adversarial attacks (e.g., prompt injection~\cite{greshake2023not,liu2023prompt}, backdoor attacks~\cite{liu2024demystifying}). SE plays a critical role in establishing systematic security mechanisms, including secure access control, privacy-preserving techniques (e.g., differential privacy~\cite{charles2024fine}, federated learning~\cite{ye2024openfedllm}), and trusted execution environments (TEE)~\cite{mohan2024securing}. For instance, Azure OpenAI Service~\cite{azure_openai_service}, as an AI service provider, implements strict role-based access control (RBAC) to ensure that users can only access authorized data and functionalities. Concurrently, research efforts are exploring the application of differential privacy in data processing, as demonstrated by Google's work~\cite{kurakin2023harnessing}, to prevent sensitive training data from being exposed during inference. Moreover, in multi-LLMs deployment scenarios, ensuring secure inference environments through containerization and sandboxing techniques (e.g., Intel SGX~\cite{intel_sgx}) is essential. These approaches isolate user inputs during inference, preventing unauthorized access and adversarial exploitation, thereby significantly enhancing the security and trustworthiness of LLM.

\section{Requirements Engineering}\label{sec:re}
\begin{figure}[htbp]  
    \centering     
    \includegraphics[width = \textwidth]{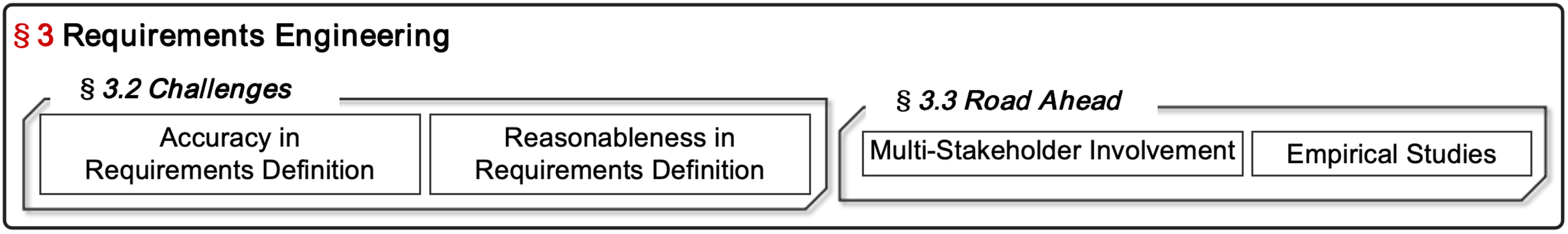}  
    \caption{Challenges and Road Ahead in \S\ref{sec:re} Requirements Engineering.}
    \label{fig:re_road_ahead}

\end{figure}

From this chapter onward, we analyze the research status of each phase in the LLM development lifecycle, identify the challenges, and propose potential future directions. For RE of LLM, the challenges and potential future directions are shown in \autoref{fig:re_road_ahead}.



\subsection{Research Status}
To the best of our knowledge, research on RE for LLMs remains relatively limited. Due to the strong natural language processing (NLP) capabilities of LLMs, existing studies primarily focus on leveraging LLMs to support RE tasks, while comparatively fewer efforts investigate RE methodologies for LLM development itself. This imbalance mirrors a similar trend observed in the broader AI domain. As noted by Ahmad \ea{}~\cite{ahmad2023requirements}, between 2011 and 2021, only approximately 43 publications explicitly addressed RE for AI, whereas a substantially larger work explored the use of AI techniques to enhance RE processes.

However, this imbalance does not imply that RE for LLMs is insignificant. As LLMs are increasingly applied across diverse domains, they encounter distinct requirements based on the specific demands of different scenarios. It is essential to thoroughly understand these requirements and tailor the development of LLMs accordingly. Fischer \ea{}~\cite{fischer2024multi} fine-tuned a model based on the requirements of investigative intelligence, yet their understanding of user requirements was mainly derived from prior research. In contrast, Solomon \ea{}~\cite{solomon2024requirements} conducted an RE study to investigate the use of LLMs in digital inquiry processes aimed at enhancing healthcare applications. They proposed a generalizable RE methodology for LLMs that incorporates both qualitative and quantitative analyses. The qualitative analysis involves studying the target population's background, including cultural and linguistic factors, while the quantitative analysis utilizes techniques such as word embeddings and network analysis to construct a semantic framework for the model. Beyond user research methods, Hassani \ea{}~\cite{hassani2025empirical} addressed the requirements of a food company by fine-tuning LLMs to enhance their ability to classify legal texts related to food safety, incorporating both food safety system and software requirements. Additionally, Sjostrom \ea{}~\cite{sjostrom2024meta} proposed meta-requirements for LLM-based knowledge retrieval tools. Although the methodologies proposed in these studies are domain-specific and lack general applicability, they underscore the critical role of RE in LLM development: through rigorous RE, developers can identify the specific functionalities and performance requirements needed, ultimately enabling the customization of powerful, task-specific models.

In addition to the domain-specific requirements for LLMs discussed above, several general requirements recur across diverse scenarios. These include requirements related to dataset quality, energy efficiency, user preferences, and model interpretability. Dataset quality is a key factor influencing the performance of LLMs~\cite{liu2024datasets}. Despite its importance, there remains no clear consensus on which specific quality metrics are most relevant or how they should be applied. As a result, researchers continue to face difficulties in consistently evaluating and comparing dataset quality across different tasks and domains. In resource-constrained devices such as edge devices (as discussed in \S\ref{sec:edge_deployment}), LLMs may meet requirements related to energy consumption and computational efficiency. User preference requirements are also increasingly prominent, as policies, cultural values, and ethical standards vary significantly across regions. These contextual factors influence an acceptable LLM response. Finally, explainability remains a key concern. Since LLMs function primarily as black boxes, it is often unclear how they arrive at specific outputs. This lack of transparency raises important questions about the reliability, accuracy, and trustworthiness of their responses~\cite{zhao2024explainability}.


\subsection{Challenges}
Even before the emergence of LLMs, the impressive capabilities of AI had already given rise to misconceptions that AI could address all problems~\cite{ahmad2023requirements}. With the advent of LLMs, these expectations have grown even further, leading to an increased demand for functional requirements (FRs). Concurrently, the inherent complexity of LLMs has given rise to a diverse array of non-functional requirements (NFRs), such as interpretability, robustness, and efficiency. We categorize the key challenges of RE for LLMs into two dimensions: the accuracy and the reasonableness of requirement definitions.

\textbf{Accuracy in requirements definition.}  
Clearly defining requirements, whether FRs or NFRs, remains a challenging task. For example, Hassani \ea{}~\cite{hassani2025empirical} found it difficult to determine which food safety regulations applied to software requirements, as these laws were not formulated initially with digital systems or AI integration in mind. Similarly, when defining NFRs such as creativity, conceptual ambiguity becomes a significant obstacle. Questions such as “What constitutes creativity in the context of LLMs?”, “Which domains should it be evaluated in?” and “How can it be measured objectively?”  are still far from being resolved~\cite{franceschelli2024creativity}.

\textbf{Reasonableness in requirements definition.}  
Beyond accuracy, it is equally essential to ensure that requirements are reasonable and achievable. Conflicting requirements or unrealistic expectations often necessitate trade-offs to formulate practical and balanced specifications. For instance, in edge deployment scenarios involving resource-constrained devices, a certain degree of performance degradation is inevitable. So defining acceptable trade-offs, such as how much accuracy can be sacrificed while maintaining robustness, or how much memory and computing resources the model is allowed to use, presents a complex challenge. Output consistency is another challenge. Due to the inherently probabilistic behavior of LLMs, it is often unrealistic to expect identical outputs for the same input, which calls for more flexible and context-aware approaches to defining such requirements. Together, these challenges highlight the need for refined RE methodologies tailored to LLMs—methods that can more accurately and practically define both FRs and NFRs.

\subsection{Road Ahead}
To address the challenges outlined above, we propose two feasible research directions. First, to improve the accuracy of requirements definition, we recommend adopting a \textbf{multi-stakeholder involvement} strategy, which engages users, developers, and domain experts in a collaborative process to reach consensus on requirements. This approach has been successfully applied in several specific scenarios. For instance, Solomon \ea{}~\cite{solomon2024requirements} collaborated with medical professionals to validate the accuracy of their analytical framework and established general principles for requirement definition in digital healthcare contexts. Similarly, Chakrabarty \ea{}~\cite{chakrabarty2024art} recruited writers, volunteers, and domain experts to define LLM creativity, incorporating expert insights alongside standardized creativity assessment methods such as the Torrance Tests of Creative Thinking (TTCT)~\cite{torrance1966torrance}. However, these examples mainly focus on specific, well-defined user groups. In scenarios involving international or cross-domain user bases, the effectiveness and scalability of such multi-stakeholder approaches become limited. Therefore, a promising future direction is to investigate how to more systematically and scientifically incorporate diverse stakeholder perspectives when defining requirements across broader, more heterogeneous application environments. 

Second, to address the challenge of defining reasonable requirements, we advocate for increased emphasis on \textbf{empirical studies}. Such studies enable a systematic understanding of the trade-offs involved in LLM deployment by evaluating model performance across diverse scenarios, which facilitates the establishment of practical, evidence-based requirement boundaries. For example, Huang \ea{}~\cite{huang2024empirical} conducted an extensive evaluation of LLaMA3 quantization across 1-8 bits settings, yielding empirical insights into the trade-offs between model performance and memory efficiency. Nevertheless, such empirical efforts are still relatively limited, highlighting the need for further research in this direction. 


\section{Dataset Construction }\label{sec:datasets}
%

\begin{figure}[htbp]  
    \centering     
    \includegraphics[width = \textwidth]{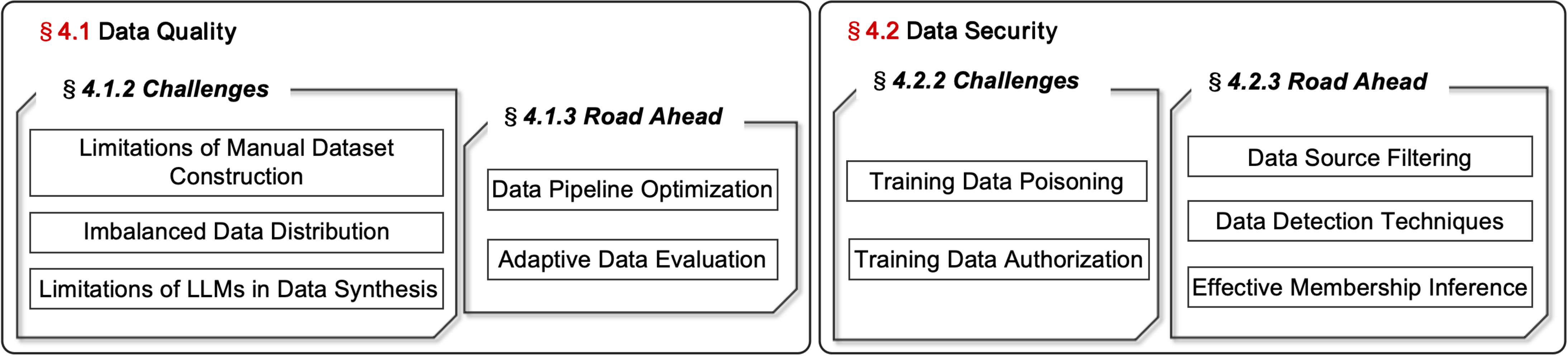}  
    \caption{Challenges and Road Ahead in  \S\ref{sec:datasets} Dataset Construction.}
    \label{fig:datasets_road_ahead}

\end{figure}

In the development of LLMs, training datasets, encompassing both pre-training and fine-tuning corpora, play a crucial role. In this section, we analyze datasets primarily from two critical dimensions: data quality and data security. These aspects not only influence model performance and generalization but also raise technical and ethical challenges. As shown in \autoref{fig:datasets_road_ahead}, by focusing on our discussion around these two dimensions, we aim to provide a comprehensive perspective on the construction and utilization of datasets in LLM development.

\subsection{Data Quality}\label{sec:data-quality}

\subsubsection{Research Status}

Data quality directly influences the diversity, relevance, and accuracy of datasets, which are critical factors in improving LLM performance for specific tasks~\cite{liu2024datasets}. Feng \ea{}~\cite{feng2024pre} demonstrated a positive correlation between the frequency of causal relationships in pretraining corpora and LLM performance in causal discovery tasks. Similarly, Rao \ea{}~\cite{rao2023cat} proposed a pre-training approach that leverages the mapping between code and test files to enhance the relevance of training data, thereby improving LLM-generated test cases. As a result, obtaining high-quality datasets has become a key focus of research.

According to the construction method, we broadly categorize current approaches to improving data quality into two main strategies: \textit{(a)} manual data labeling and rule-based selection, and \textit{(b)} LLM-assisted data construction.  
\textbf{Manual methods} typically yield high-quality datasets~\cite{conover2023free,singh2024aya,xue2023instruction,li2023starcoder} but are labor-intensive and often result in relatively small datasets.  
While \textbf{LLM-assisted methods} involve using LLMs to label, synthesize, or filter high-quality data automatically. Due to their strong performance, LLMs have been extensively employed for data construction~\cite{liu2024automatic,chung2023increasing,wang2022self,nayak2024learning} to facilitate large-scale dataset generation through dedicated pipelines or agents. Correspondingly, there are two primary LLM-based data generation methods: \textbf{reference-based methods} and \textbf{collaborative LLM methods}. Reference-based methods typically leverage high-quality seed datasets, strong baseline models, or external knowledge sources as references to guide the generation of higher-quality datasets. For instance, Gao \ea{}~\cite{gao2024learning} introduced a teacher-student framework where an LLM extracts high-quality samples from unlabeled data by comparing them against a reference dataset.  
Collaborative LLM methods involve multiple LLMs working together to identify high-quality datasets. For example, Liu \ea{}~\cite{liu2024coachlm} fine-tuned datasets using human expert-created instructions to produce richer and more precise instruction datasets. Similarly, Huang \ea{}~\cite{huang2024effi} employed LLMs to optimize fine-tuning datasets for improved code generation efficiency. Despite their advantages, these methods share a common limitation: their effectiveness is inherently constrained by the performance of the LLM itself. If the LLM's capability is suboptimal, the quality of the generated datasets is affected~\cite{yu2024large}.

As an essential component of data quality, data diversity has a significant impact on model generalization and robustness. Thus, it has gained people's attention. Zhou \ea{}~\cite{zhou2023devil} demonstrated the presence of long-tail effects in datasets, where LLMs exhibit lower performance on rare data categories. Traditional approaches, such as Focal Loss~\cite{ross2017focal} and Learning-to-Rank (LTR)~\cite{alshammari2022long}, have been proven ineffective in mitigating these issues for LLMs. 

Therefore, approaches for enhancing data diversity have been widely studied in recent years, which can be broadly classified into two categories: \textit{(a)} preserving diverse samples during data selection and \textit{(b)} employing data synthesis techniques. Although real-world data is inherently diverse~\cite{penedo2023refinedweb}, its imbalanced distribution poses significant challenges~\cite{longpre2023pretrainer}, such as minor languages remaining persistently underrepresented, which causes the related corpus to be significantly rare~\cite{masala2024vorbe, Khan_2024}. Due to the high cost of manual efforts, researchers often rely on LLMs to select diverse data samples automatically. However, their capability to directly assess data diversity remains limited~\cite{pang2024improving}. Consequently, researchers have turned to data synthesis techniques to enrich underrepresented categories~\cite{long2024llms}. Also, due to the high costs associated with manual data synthesis~\cite{liu2024datasets}, recent efforts have focused on leveraging LLMs for automated data generation~\cite{xu2024ds2absadualstreamdatasynthesis,whitehouse2023llm,chung2023increasing}. Yuan \ea{}~\cite{yuan2021synthbio} applied this method to synthesize biographical texts, reducing biases associated with occupation and improving dataset balance. Additionally, LLMs can serve as translators, converting widely available data from natural languages~\cite{zhu2024dynamic}, programming languages~\cite{liu2024your,cassano2022multipl}, or multimedia formats~\cite{lupidi2024source2synth} into less common languages, programming paradigms, or textual information. This approach generates rare data types at scale while minimizing human effort. However, due to inherent biases in LLMs, it may inadvertently introduce biases or errors~\cite{duan2024large,xu2024pride}.


\subsubsection{Challenges}

The findings discussed above highlight several \textbf{challenges} in improving data quality, which can be categorized into three key aspects:  

\textbf{Limitations of manual dataset construction.}  
Manual data creation is labor-intensive and inherently constrained in scale. A significant drawback is its time-consuming nature, as tasks such as dataset cleaning, labeling, and ensuring balanced data distribution require extensive human effort. Although pipelines and custom rules can assist in data collection and filtering, we still can not achieve full automation to process data. Furthermore, it is impossible to mitigate biases and achieve a well-balanced dataset distribution only through rule-based automation. As a result, the dependence on manual processes limits the speed and scalability of dataset development.  

\textbf{Imbalanced data distribution.}  
Data imbalance is common in both real-world distributions and training dataset distributions. Different languages~\cite{longpre2023pretrainer, Khan_2024}, geographic regions, time periods~\cite{masala2024vorbe}, and data sources contribute to this imbalance, which complicates data acquisition and processing. As a result, imbalanced datasets introduce challenges:  
\textit{(a)} The long-tail effect~\cite{zhou2023devil}, where LLMs perform poorly on underrepresented data categories.  
\textit{(b)} Inherent biases.  When sources like social media dominate datasets, they often carry inherent biases~\cite{lin2024investigating}. These biases can harm model generalization~\cite{li2025understanding} and raise ethical issues~\cite{jiao2024navigating}.

\textbf{Limitations of LLMs in data synthesis.}  
Despite their advancements, LLMs exhibit inherent limitations in improving data quality and generating synthetic data~\cite{yu2024large}. Since many of their abilities are still not comparable to human experts, fully relying on automation for data filtering, labeling, and evaluation is often ineffective. For instance, Pang \ea~\cite{pang2024improving} found that LLMs struggle with accurately assessing data diversity. While LLMs have shown promise in data synthesis, their inherent biases can result in imbalanced synthetic data distributions~\cite{duan2024large,xu2024pride}, further complicating the challenge of dataset construction.

\subsubsection{Road Ahead}

To enhance data quality, we propose two potential research directions. \textbf{Data pipeline optimization} is a promising direction. Specialized data pipelines or agents for collecting and processing LLM training data have been explored by Ostendorff \ea{}~\cite{ostendorff2024llm}. Furthermore, such architectures should incorporate automated tools to efficiently filter or generate target data based on user configurations. In addition to essential automation tools, LLMs can assist in this process. However, given the limitations of a single model, leveraging multi-model collaboration and multimodal data transformation can help overcome these constraints. Furthermore, incorporating human expert feedback can enhance data quality while mitigating the high costs associated with manual dataset construction.  

At the same time, \textbf{adaptive data evaluation} can further improve data diversity by establishing robust evaluation mechanisms. Potential approaches include:  
\textit{(a) Dynamic long-tail adaptation}, which adjusts data generation in real time based on distribution patterns to prevent imbalances.  We can incorporate models such as support vector machines (SVMs) or clustering algorithms to automatically classify data and infer its distribution. However, data distribution detection faces two main challenges: how to define suitable classification criteria and how to determine the categories to be used.
\textit{(b) Multimodal contextual assessment}, which utilizes advanced multimodal translation techniques to generate data across different modalities based on evaluation results. This method facilitates cross-modal data transformation to enhance data diversity and improve overall data quality. 
\textit{(c) Bias-aware diversity scoring frameworks}. 
Although LLM-as-a-judger has become increasingly popular, it remains unsuitable for reliably identifying bias. In the future, we could build a clear and practical framework for evaluating bias, including clear categories of bias and corresponding assessment metrics. Within this framework, LLMs could serve as auxiliary tools to match outputs with relevant bias categories or similar cases. By analyzing the distribution of outputs within the same category, it would then be possible to determine whether a particular LLM response exhibits bias more accurately.


\subsection{Data Security}
\subsubsection{Research Status}

As discussed in \S\ref{sec:data-quality}, datasets may contain biased or harmful content, making LLMs susceptible to generating incorrect or unsafe outputs~\cite{pathmanathan2024poisoning}. This issue poses significant risks in critical domains such as healthcare~\cite{alber2025medical}. One major cause of this problem is data poisoning~\cite{ji2024beavertails,jiang2024turning,xu2025shadowcast}, where adversarially manipulated data is put into training datasets, leading to unintended behaviors in LLMs. While existing studies~\cite {wang2022revise,pagano2023bias} have explored methods to mitigate dataset bias, more advanced techniques are still needed to identify increasingly complex bias patterns.  

Beyond biased data, the widespread adoption of LLMs for code generation has raised concerns regarding malicious code embedded in training datasets, which can lead LLMs to produce security vulnerabilities~\cite{mohsin2024can,cotroneo2024vulnerabilities}. Yan \ea{}~\cite{yan2024llm} and Liu \ea{}~\cite{liu2024eatvul} demonstrated that LLMs can synthesize vulnerable code capable of evading traditional static analysis tools as well as LLM-based vulnerability detection mechanisms. Although researchers have proposed countermeasures at different stages, including during code generation~\cite{nazzal2024promsec,li2024cosec} and post-generation analysis~\cite{xue2024poster}, the issue of preventing malicious code at the dataset level has received limited attention from researchers.  

Additionally, since many LLM training datasets are not publicly disclosed, concerns have emerged regarding the use of unauthorized data, raising complex intellectual property and legal issues~\cite{zhou2023don,xu2024benchmark}. To address the issue of training data authorization, researchers have explored membership inference techniques~\cite{shi2023detecting,maini2024llm,chen2024catch,dealcala2024my} to determine whether a specific dataset was used in model training. For instance, Shi \ea{}~\cite{shi2023detecting} proposed a method to detect whether a dataset was incorporated into an LLM’s training data by analyzing whether it was publicly released after the model’s training period. However, this approach does not infer internal relationships within datasets. To overcome this limitation, Maini \ea{}~\cite{maini2024llm} trained a linear model and used information scores to determine if a dataset was part of the model’s training set. However, this method is still not precise enough. Another technique to address data authorization is digital watermarking~\cite{guo2025zeromark,cui2025ft,pan2025can,wegerhoff2024datadetective}, which embeds markers to trace unauthorized usage and ensure data provenance. However, due to the current limitations of watermarking technology, such as requiring black box access rights~\cite{wang2025sleepermark} or being vulnerable to attacks~\cite{pang2024attacking}, it still needs to be further improved. 


\subsubsection{Challenges}

The aforementioned research findings highlight two key challenges in ensuring the security of training data: \textbf{training data poisoning} and \textbf{training data authorization}. Each aspect presents unique risks and requires targeted mitigation strategies to address them.

\textbf{Training data poisoning.}  
LLM training relies on large-scale datasets, making it highly susceptible to data poisoning attacks. Adversaries can deliberately inject malicious information, misleading content, or backdoor data into training sets to manipulate model behavior or even exert control over its generated outputs. These attacks include backdoor attacks, knowledge contamination, ethical and security pollution, and steganographic attacks. Even if a training dataset contains only a small fraction of malicious code, the trained LLM may still generate vulnerable code with high probability~\cite{pathmanathan2024poisoning}. However, defending against data poisoning remains highly challenging due to LLMs’ reliance on extensive, unverifiable datasets. While techniques such as poisoned data detection~\cite{baracaldo2017mitigating} and secure prompt engineering~\cite{yang2025guardt2i,zhou2025robust,zhang2024adversarial} have been developed to mitigate the impact of data poisoning, limited research has focused on systematically detecting and filtering malicious data within training datasets~\cite{das2025security}.

\textbf{Training data authorization.}  
 Training data is often sourced from open datasets, web-scraped content, and user-provided data. However, not all data is explicitly authorized for commercial use or model training, posing legal risks related to copyright infringement, privacy compliance, and platform policies. Currently, membership inference and digital watermarking techniques are the primary methods used to address data authorization concerns.  However, these techniques have their limitations. For example, some methods rely on disclosure time information or require black-box access to the model, which makes data verification less accurate and limits their applicability. 


\subsubsection{Road Ahead}\label{sec:data_security_ra}

Currently, no fully effective solution exists for mitigating training data poisoning~\cite{das2025security}, but several promising directions are worth further exploration.  One potential approach is \textbf{data source filtering}, which involves establishing trusted data sources and exclusively collecting data from these sources. When combined with data provenance techniques to track the origin and modification history of data, this method enhances the traceability of datasets. However, it faces notable challenges, including ensuring the reliability of trusted sources and the limited availability of high-quality training data.  Another promising direction is the advancement of \textbf{data detection techniques}. While some studies~\cite{jiang2024turning,cotroneo2024vulnerabilities,fu2024poisonbench} have been conducted, more precise detection methods are required to identify malicious data, biased content, and trigger patterns. To address this, a data auditing and risk assessment platform could be developed, integrating automated tools for real-time or periodic security audits to identify data quality and security issues. Given LLMs' strong analytical capabilities and their successful applications across various domains~\cite{kaddour2023challenges}, it is viable to leverage them for finer-grained detection. Specifically, a real-time anomaly detection system powered by LLM-assisted analysis could be designed to automatically identify and block anomalous data before it enters the training pipeline, thereby minimizing its potential negative impact.  

To address data authorization concerns, integrating data provenance with blockchain technology represents a potential solution. However, storing large-scale datasets on-chain remains a significant challenge. Even if only dataset hashes are recorded on-chain, this approach becomes significantly less effective when there is little to no information available about the datasets used to train the LLMs. Thus, \textbf{effective membership inference} is essential. Although recent studies~\cite{maini2025llm,meeus2024did} have made notable progress, they fall short of addressing deeper, structural challenges that remain at the core of the problem. The method proposed by Maini \ea{}~\cite{maini2025llm} requires independently and identically distributed (IID) datasets, while the approach in Meeus \ea{}~\cite{meeus2024did} struggles with small-scale inference and its accuracy remains limited. Future research should focus on enhancing membership inference methods to achieve higher accuracy and finer granularity.

\section{Development and Enhancement}\label{sec:development_enhancement}

\begin{figure}[htbp]  
    \centering     
    \includegraphics[width = \textwidth]{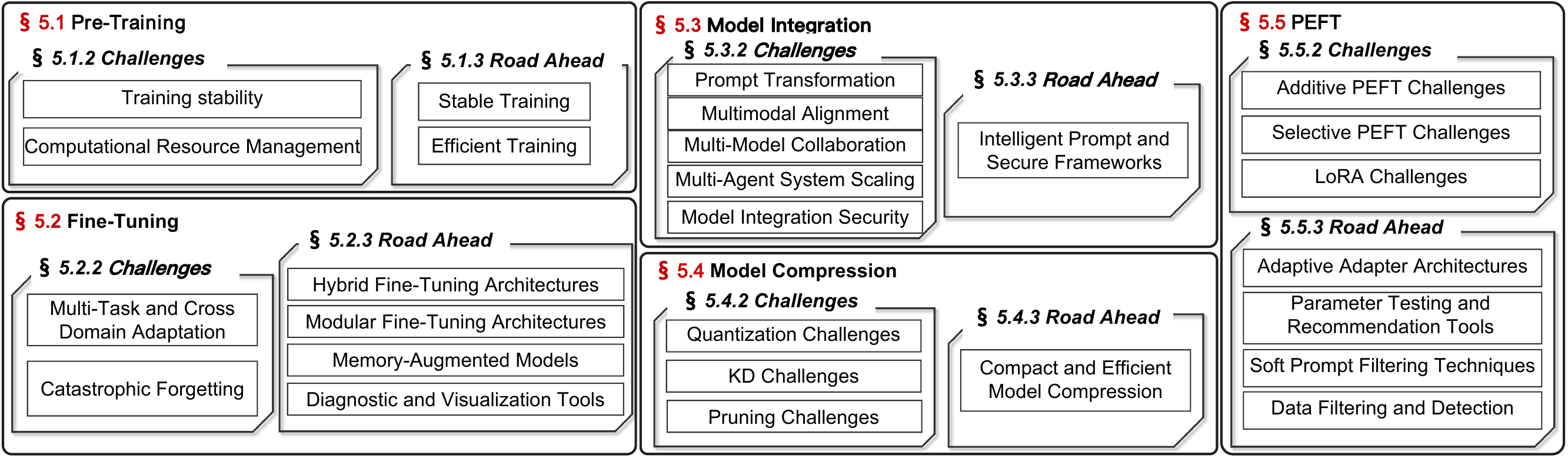}  
    \caption{Challenges and Road Ahead in \S\ref{sec:development_enhancement} Development and Enhancement.}
    \label{fig:development_road_ahead}

\end{figure}

The development of LLMs has revolutionized the field of AI. However, it also introduces unique challenges from an SE perspective. Model enhancement aims to improve a model's capabilities, performance, and reliability. Notably, since models can be enhanced during development through fine-tuning or continual training, a strong coupling exists between the development and enhancement phases.  Due to this interdependence, we explore the critical challenges associated with both LLM development and enhancement, focusing on three primary phases of model development: pre-training, fine-tuning, and model integration, as shown in \autoref{fig:development_road_ahead}. We first analyze existing limitations and emerging solutions, followed by a discussion of key techniques: model compression and PEFT.

\subsection{Pre-Training}

\subsubsection{Research Status}\label{sec:pre-training-rs}

Pre-training is the foundation of LLM development, yet it faces challenges in terms of scale, resource demands, and engineering complexity. Ensuring and enhancing the effectiveness of pre-training has been a long-standing research focus, with efforts directed towards optimizing training datasets, refining training methodologies, and improving computational efficiency~\cite{lin2024vila,tirumala2023d4}. SE plays a critical role in advancing LLM pre-training, contributing to areas such as automating training pipelines~\cite{he2021pipetransformer,luo2024autom3l}, optimizing training architectures~\cite{narayanan2021efficient}, enhancing stability and security~\cite{mo2024large,wang2025scale,xie2024optimization,chen2025efficient}, and reducing energy consumption~\cite{arfeen2024pipefill}.  

As models continue to grow in size and complexity, maintaining training stability has become an increasingly critical challenge~\cite{tang2023improving}, particularly in ensuring smooth convergence. Existing research primarily focuses on mitigating gradient explosion~\cite{wang2025scale} and gradient vanishing~\cite{mo2024large}, as well as optimizing learning rates~\cite{xie2024optimization,chen2025efficient}, all of which aim to regulate parameter distributions and transformations during training to enhance stability. For instance, Chung \ea{}~\cite{chung2024stable} controlled output layer embedding variance to prevent gradient explosion, while Agarwal \ea{}~\cite{agarwal2024policy} and Woo \ea{}~\cite{woo2025dropbp} improved stability by selectively discarding specific backpropagation steps. Nishida \ea{}~\cite{nishida2024initialization} attributed loss spikes and convergence failures to uneven parameter distributions and introduced weight scaling as reparameterization (WeSaR) to normalize parameter norms for stable training. Similarly, Wortsman \ea{}~\cite{wortsman2023stable} proposed a hybrid AdamW-Adafactor optimizer to mitigate loss spikes. Moreover, parameter precision also impacts training stability. DeepSeek-V3~\cite{liu2024deepseek} preserved the original precision of specific model components to ensure stable training.

The exponential growth in model sizes has further intensified computational resource constraints~\cite{sani2024future,bai2024beyond,xu2025resource,wu2024evolutionary,shen2025search}. Training SOTA LLMs requires extensive computational infrastructure, which costs potentially reaching millions of dollars per training run. One commonly adopted approach to alleviate this challenge is model compression~\cite{zhu2024survey}, which reduces model size to lower resource demands. However, these methods often result in some degree of performance degradation. We will discuss model compression techniques in detail in \S\ref{sec:model_compression}.

\subsubsection{Challenges} 

The research mentioned above highlights two key challenges in pre-training:  

\textbf{Training stability.}  
Although various techniques have been proposed to improve training stability, most approaches rely on heuristics rather than systematic frameworks. For instance, learning rate (LR) warm-up is commonly employed to mitigate gradient explosion and enhance stability. However, there is currently no general model for evaluating its effectiveness across different LLM architectures and training setups~\cite{xiong2020layer}. In practice, parameter settings are often chosen through trial and error, based on the needs of individual cases~\cite {ibrahim2024simple,guo2024efficient}. Moreover, as model sizes continue to increase, the training process becomes increasingly complex and difficult to regulate; therefore, it is necessary to conduct systematic research into training stability.  

\textbf{Computational resource management.}  
As LLMs continue to scale, managing computational resources has become a major challenge—training a single large model can cost millions of dollars in hardware, energy, and time~\cite{xu2024survey,duan2024efficient}. This economic burden limits the capability of smaller research teams and companies with constrained resources to develop proprietary LLMs, limiting opportunities for smaller research teams to innovate and allowing a few major proprietary models to maintain their leading positions in the field. Lowering training costs would make it easier for smaller labs and open-source communities to build and improve their models, helping to reduce the dominance of a few major players. In the industry, the cost-efficient training approach adopted in DeepSeek-V3~\cite{liu2024deepseek} has already garnered significant attention, further facilitating the development of more accessible and cost-effective models.

\subsubsection{Road Ahead}

To address these challenges, potential solutions can be explored from two key directions: \textbf{stable training} and \textbf{efficient training}.  

\textbf{Stable training.}  
One of the primary causes of training failures in LLMs lies in internal parameter updates. As discussed in \S\ref{sec:pre-training-rs}, most existing approaches rely on heuristic adjustments rather than rigorous theoretical foundations. Zucchet \ea{}~\cite{zucchet2025recurrent} explored optimization challenges in RNNs and identified fundamental causes of gradient explosion. Similarly, \textbf{achieving a deeper theoretical understanding of training dynamics in LLMs} is essential for identifying key instability factors, such as the underlying triggers of loss spikes.  Beyond theoretical advancements, the development of \textbf{real-time monitoring and analysis tools} for LLM training could enable the detection and prediction of anomalous behaviors by tracking stability-related metrics. Additionally, it is important to design user-friendly visualization tools and interactive interfaces that help researchers interpret model behavior and monitor the training process more effectively.

\textbf{Efficient training.}  
Due to the high computational costs associated with pre-training, improving LLM efficiency is a crucial research direction. 
GPUs used for model training are extremely costly, which makes it essential to have tools that can monitor usage in real time and adjust workloads to avoid waste. \textbf{These tools should minimize GPU usage without increasing training time or compromising model performance}, thereby significantly reducing overall computational costs. Additionally, model growth techniques from ML in which smaller models are leveraged to accelerate the training of larger ones hold promise for improving efficiency. While this approach has not yet been widely adopted in LLM pre-training, Du \ea{}~\cite{du2025stacking} conducted an empirical study providing insights into its potential application. Therefore, \textbf{it is considered valuable to explore the application of model growth techniques in LLM development in the future.}

\subsection{Fine-Tuning} 

\subsubsection{Research Status}

Fine-tuning enables pre-trained LLMs to adapt to specific tasks while balancing adaptation, knowledge retention, computational efficiency, and deployment constraints. However, it presents several challenges, including PEFT (will be discussed in detail in \S\ref{sec:peft}), catastrophic forgetting prevention, and cross-domain generalization. 

Fine-tuning has been widely employed to enhance model performance across diverse tasks~\cite{schmirler2024fine,peters2019tune}. While fine-tuned models often exhibit significant improvements in individual tasks, effectively fine-tuning LLMs for multi-task scenarios remains a major challenge~\cite{liu2024mftcoder}. Existing approaches frequently struggle with task interference, optimal resource allocation across different objectives, and maintaining consistent performance across diverse domains~\cite{xia2025efficient,tian2025hydralora,xin2024beyond,li2024unifiedmllm}. To address these challenges, researchers have explored techniques such as LoRA~\cite{luo2024zero,agiza2024mtlora} and MoE~\cite{zhu2024uni,yang2024multi,kim2021scalable}, further enhanced by optimization techniques~\cite{wang2024arithmetic,li2024s} and resource allocation strategies~\cite{liu2024moe,oh2024exegpt}. However, one major challenge is improving task-specific performance while still preserving the model’s general ability to work across different domains.  

Additionally, fine-tuning introduces the risk of catastrophic forgetting, wherein models lose previously acquired knowledge during adaptation. Recent studies suggest that what appears to be forgetting may be caused by the model becoming less aligned with the task, rather than truly losing the knowledge it learned~\cite{zheng2025spurious}. However, the underlying causes are still unclear~\cite{li2024revisiting}. This problem becomes even more noticeable when the model needs to be updated regularly~\cite{gupta2024model} or adapted to new domains~\cite{liu2024more}, as it often struggles to retain its original capabilities while learning new ones.

LoRA has demonstrated potential in mitigating catastrophic forgetting and has achieved notable success~\cite{ren2024analyzing,ni2023forgetting,dou2024loramoe}. However, recent studies indicate that LoRA still struggles with certain limitations, such as instruction-following constraints~\cite{jiang2024interpretable} and scaling challenges~\cite{kalajdzievski2024scaling}. Continual learning approaches have been proposed to enable iterative knowledge acquisition while preventing forgetting~\cite{fawi2024curlora,srivastava2024improving}. Nevertheless, research has shown that continual learning can lead to significant performance degradation after repeated training cycles~\cite{li2024examining,huang2024mitigating,li2024moe}. Furthermore, these approaches are not universally applicable across different modalities in multimodal LLMs~\cite{zhai2024investigating}, underscoring the need for modality-specific fine-tuning solutions.

\subsubsection{Challenges}

Beyond PEFT, which will be discussed in \S\ref{sec:peft}, we identify two key challenges in fine-tuning.  

\textbf{Multi-task and cross-domain adaptation.}  
Fine-tuning LLMs for multiple tasks or domains simultaneously presents several challenges~\cite{liu2024mftcoder}, including task interference, optimal resource allocation, and maintaining consistent performance across diverse domains. While existing approaches such as LoRA and MoE have demonstrated effectiveness in mitigating these issues, there are still further improvements needed in areas such as training efficiency~\cite{wang2024malora}, model generalization~\cite{li2024unifiedmllm}, and system overhead during task switching~\cite{xia2025efficient}.  

\textbf{Catastrophic forgetting.}  
The underlying mechanisms behind catastrophic forgetting in LLMs remain largely unexplored, and no highly effective solutions have been developed to address this issue comprehensively. Although existing techniques, such as LoRA and continual learning, can help mitigate forgetting, they often come at the expense of performance degradation or the loss of other learned knowledge. Furthermore, these approaches are ineffective in multimodal models, which highlights the importance of developing fine-tuning methods tailored to each modality.

\subsubsection{Road Ahead}

The challenges of catastrophic forgetting and multi-task, multi-domain adaptation underscore the lack of universality in current fine-tuning methods, which have yet to achieve the goal of adapting to diverse tasks~\cite{peters2019tune}. Future fine-tuning approaches should aim to enable both effective multi-task adaptation and mitigate catastrophic forgetting.  

For \textbf{multi-task and multi-domain adaptation}, a promising direction is to integrate various fine-tuning techniques, such as LoRA, with MoE, quantization, and resource optimization strategies, forming a \textbf{hybrid fine-tuning architecture}. Specifically, designing an architecture capable of real-time monitoring of task interference could enable the rapid detection of adverse transfer effects, allowing for automated adjustments in task allocation and fine-tuning strategies. For instance, dynamically allocating additional resources to tasks with higher interference could mitigate performance degradation. This approach has the potential to minimize or eliminate task interference while preserving model generalization and maintaining inference efficiency.  In addition to combining different fine-tuning methods, a \textbf{modular fine-tuning architecture} could be designed to make it easier for LLMser to switch between, plug in, or combine modules tailored to specific tasks or domains. Such an approach would reduce the coupling between fine-tuning techniques, thereby lowering maintenance costs and improving adaptability. Notably, this idea aligns with the design principles of industrial MoE and LoRA modules, which primarily focus on enhancing LLM performance. Recent studies on MoE and LoRA~\cite{li2024locmoe,kang2024self,xue2024openmoe} have begun exploring intelligent scheduling mechanisms to support multi-task adaptation further.  

For \textbf{catastrophic forgetting mitigation}, an essential research direction is the development of comprehensive methods to assess the extent and content of knowledge forgotten by LLMs. Such evaluations are crucial for facilitating targeted recovery of forgotten information and improving model retention strategies. Additionally, knowledge retention techniques tailored for multimodal models are urgently needed, as traditional PEFT and continual learning methods have demonstrated limited effectiveness in this context. Future research could explore \textbf{memory-augmented models} that leverage external memory mechanisms, such as knowledge graphs and vector databases, to reduce reliance on parameter-based memory. Furthermore, inspired by LR warm-up strategies, progressive adaptation~\cite{hou2024progressive} could be investigated as a gradual fine-tuning approach to prevent large learning rates from causing gradient explosion or vanishing, thereby mitigating the effects of forgetting.  Another significant challenge is that catastrophic forgetting often remains undetected until post-fine-tuning evaluation, making real-time performance assessment difficult. This lack of visibility underscores the need for \textbf{diagnostic and visualization tools} specifically designed to monitor catastrophic forgetting. Future research could focus on building a diagnostic and visualization platform that helps developers track and understand how well knowledge is retained during the fine-tuning process. As Zheng \ea{}~\cite{zheng2025spurious} point out, catastrophic forgetting does not mean that the model has truly lost the knowledge. With such a platform, developers can detect signs of forgetting on time, roll back to a previous version of the model, and adjust the training strategy for retraining. In addition, automated rehearsal mechanisms or adaptive prompting strategies~\cite{huang2024mitigating} could also be integrated to proactively reduce the risk of forgetting.


\subsection{Model Integration}\label{sec:model integration}

\subsubsection{Research Status}
Unlike model development and enhancement through pre-training and fine-tuning, alternative approaches such as expanding a model’s knowledge base via Retrieval-Augmented Generation (RAG) or knowledge graph techniques, developing multimodal models, and enabling multi-model collaboration focus on integrating external models or tools with base models, known as LLM-based agents, to accomplish more complex tasks. Collectively, we refer to these approaches as \textbf{model integration}, wherein SE plays a crucial role in multiple aspects, including transforming information into prompts, coordinating interactions between models or with external tools, and optimizing routing decisions.

In LLM integration, the inclusion of RAG, knowledge graphs, additional sensors, and multimodal base models introduces a diverse and extensive range of information sources~\cite{han2024review}. Consequently, effectively transforming this information into prompts suitable for LLM task execution is important. Current research primarily focuses on preprocessing external information before generating prompts. For instance, Li \ea{}~\cite{li2024enhancing} proposed a method for summarizing contextual information before submitting it to the LLM. Similarly, in RAG, prompts are generated by combining retrieval-based methods with knowledge graphs to provide more task-relevant contextual knowledge~\cite{liu2024much,masoudifard2024leveraging,xu2024retrieval}. Moreover, tools such as EasyTool~\cite{yuan2024easytool} consolidate diverse tool-related information into unified interfaces for LLMs to process. However, these methods remain constrained by the limitations of the context window and the inherent capabilities of LLMs, preventing fundamental optimization. To address this issue, Koh \ea{}~\cite{koh2024generating} proposed a more foundational approach, mapping textual information into the embedding space of vision models to enhance image representation. Nevertheless, this approach is still limited by the constraints of the model’s embedding space. Therefore, to make real progress, new strategies are needed that go beyond the current methods.

\begin{figure}[htbp]  
    \centering     
    \includegraphics[width = \textwidth]{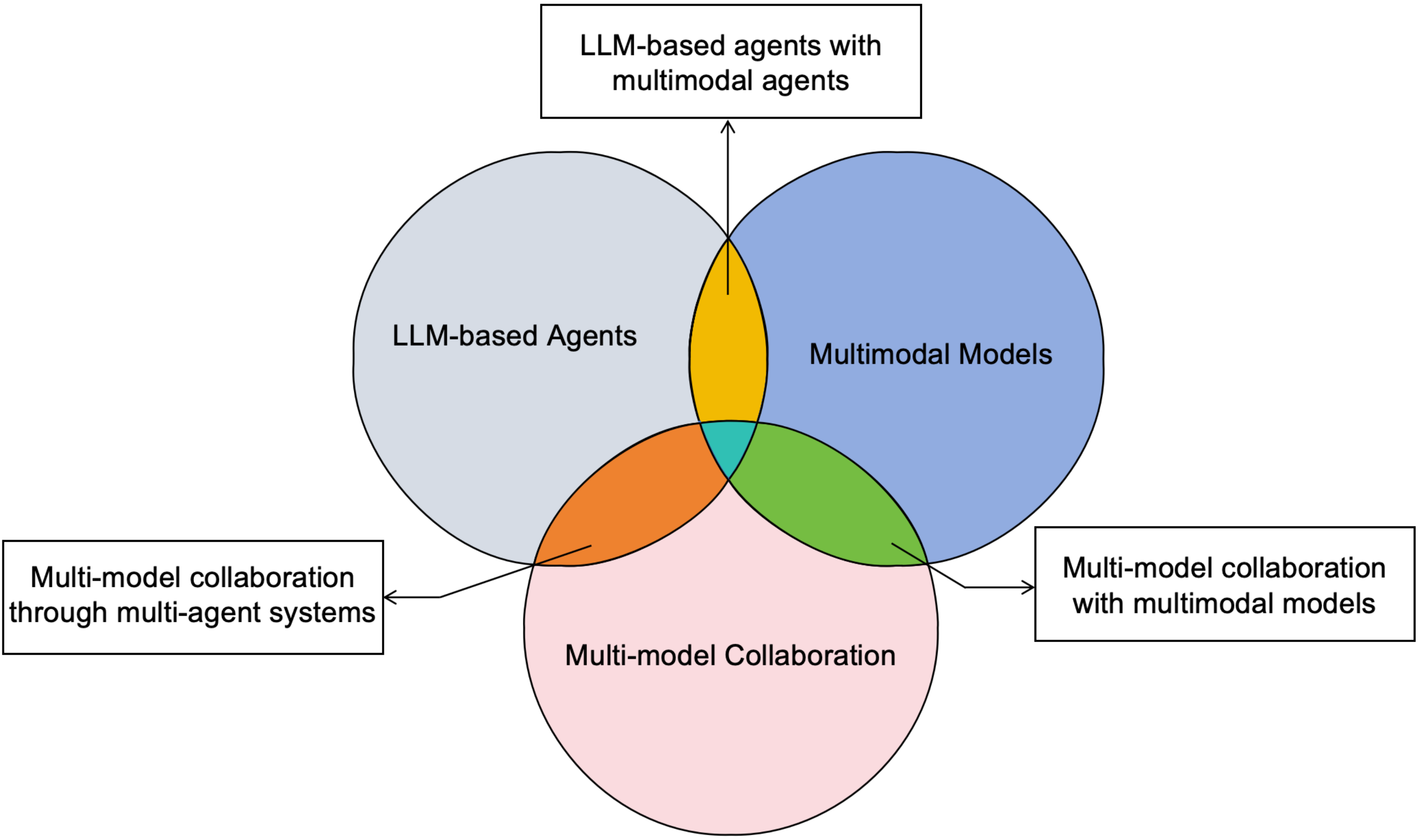}  
    \caption{Overlap between LLM-based agents, multimodal models, and multi-model collaboration.}
    \label{fig:model_integration}

\end{figure}

Given the emergence of recent protocols such as A2A, MCP, the \textbf{Agent Communication Protocol (ACP)}\cite{acp2025intro}, and the \textbf{Agent Network Protocol (ANP)}\cite{anp2025website}, which all emphasize communication and collaboration among models and tools, it is evident that the future focus of model integration lies in multi-model, multimodal processing and interactive cooperation with external tools. Accordingly, we highlight three representative forms of model integration: \textbf{multimodal models}, \textbf{multi-model collaboration}, and \textbf{LLM-based agents}. Multimodal models are capable of receiving inputs from various modalities and producing outputs across multiple modalities. Multi-model collaboration refers to leveraging cooperation, competition, or cascading among different models to accomplish complex reasoning tasks. LLM-based agents, in contrast, are built upon LLMs as the central reasoning component, and are capable of planning, decision-making, and executing tasks by interacting with external tools and knowledge sources. As illustrated in ~\autoref{fig:model_integration}, although these three paradigms differ in focus, there is overlap among them. For instance, in multi-model collaboration, complex tasks can be decomposed and distributed across models via multi-agent systems~\cite{feng2025one}. LLM-based agents can integrate multimodal models to mitigate hallucinations and enhance reasoning capabilities~\cite{jiang2024multi}, or dynamically select from a pool of multimodal models of varying types and sizes to suit different task requirements~\cite{zhang2024mm}. For clarity, we analyze these three integration paradigms separately, focusing on their unique characteristics and technical challenges, while leaving their areas of overlap outside the scope of this discussion.

For multimodal models, a fundamental challenge lies in achieving effective alignment across different modalities (e.g., text, vision, and audio)~\cite{chen2025sharegpt4v}. This challenge involves several key aspects. First, semantic consistency ensures that meaning is preserved across modalities~\cite{wang2024towards,li2024tsca}. Second, representation alignment focuses on aligning embeddings and features from different modalities to support better understanding and knowledge sharing~\cite{lei2024vit,sarkar2024xkd}. Third, cross-modal understanding aims to bridge the gaps between modalities, enabling smoother knowledge transfer and more effective interaction~\cite{huo2024c2kd}. Liu \ea{}~\cite{liu2023mitigating} highlight that failure to address these issues can result in severe performance degradation or hallucination effects. Furthermore, these challenges extend across various stages of multimodal model development, including data processing~\cite{li2024monkey,zhou2024transfusion} and pre-training strategies~\cite{li2024groundinggpt}, both of which face significant hurdles. Closing these gaps is crucial for developing multimodal models that can perform effectively across a broad range of real-world tasks.

In contrast, multi-model collaboration has been explored as a means of enhancing reasoning capabilities~\cite{patel2024embodied,liu2024two,shen2024small}. However, differences in what models can do, how they respond, and the values they reflect make it necessary to align them carefully to ensure they can work well together. Without such alignment, issues such as hallucinations may mislead reasoning processes, potentially resulting in incorrect or failed outcomes~\cite{feng2024modular,feng2024don}. Although several approaches have been proposed to mitigate these challenges, such as fine-tuning~\cite{xu2024self}, uncertainty estimation~\cite{yoffe2024debunc}, and probing other LLMs to address knowledge gaps~\cite{feng2024don}, there still lacks enough research in this area.  Another significant challenge lies in the coordination of multi-model systems, which includes managing workflows, optimizing inter-model communication, and efficiently allocating computational resources. While novel routing strategies have been introduced, such as MARS~\cite{hu2024mars}, TO-Router~\cite{stripelis2024tensoropera}, Eagle~\cite{zhao2024eagle}, and C2MAB-V~\cite{dai2024cost}, these approaches often fail to consider critical real-world constraints, such as computational resources and network bandwidth. In practice, these resource limitations introduce additional complexities~\cite{patel2024splitwise}, which will be further discussed in \S\ref{sec:model_deployment}.

Another common integration form is the \textbf{LLM-based agent}. Leveraging the powerful reasoning and decision-making capabilities of LLMs, LLM-based agents can autonomously perceive their environment, adapt to changes, and take actions when interacting with external systems, such as web services, databases, or local files. As a result, they have been widely applied in different areas~\cite{kon2025curie,wang2023intervenor, madaan2023self, huang2023agentcoder,chen2023gamegpt}. Currently, LLM-based agent frameworks typically consist of four core modules: plan, perception, memory, and action~\cite{xi2025rise, liu2024large, cheng2024exploring, luo2025large}. Specifically, the plan module is responsible for strategy formulation and execution planning, the perception module handles environmental input sensing and its transformation into representations that LLMs can understand, the memory module stores historical information to support context retrieval and ensure coherent decision-making, and the action module executes specific operations and tool invocations.


Although single-agent systems have shown impressive capabilities, they often struggle to handle complex tasks that require diverse skills, parallel processing, or coordinated decision-making~\cite{liu2024large}. To address these challenges and scale to more sophisticated problems, \textbf{multi-agent systems (MAS)} are explored, enabling division of labor and collaborative problem-solving by coordinating the efforts of multiple autonomous agents~\cite{cheng2024exploring, luo2025large}, which requires robust agent-to-agent communication and collaboration mechanisms between the agents themselves. Protocols designed for agent-to-agent communication facilitate such inter-agent collaboration, enabling agents to discover each other's capabilities, delegate tasks, and exchange information. For instance, the \textbf{A2A}~\cite{google2025a2a} specifically supports peer-like task outsourcing and dynamic interaction between autonomous agents, often within enterprise-scale workflows~\cite{ehtesham2025survey}. Beyond inter-agent communication, a fundamental requirement for both single and multi-agent systems is effective interaction with diverse external tools and resources for task execution and data retrieval. However, it is a significant challenge to standardize this agent-to-external system interaction. Additionally, existing tool integration methods are often fragmented, relying on complex manual wiring and platform-specific approaches that limit scalability and interoperability~\cite{hou2025model, cheng2024exploring, luo2025large}. To address this fragmentation and standardize AI model-to-external system interaction, the \textbf{MCP}~\cite{anthropic2025mcp} introduces a unified communication framework for LLMs to interact with external tools and resources, which simplifies tool invocation and enhances interoperability across diverse systems. Although introduced only recently, MCP has gained significant attention~\cite{singh2025survey, hou2025model}. It enables LLM-based agents to interact with external tools and systems more easily through a unified interface, facilitating the completion of complex tasks more efficiently.

Another critical issue in model integration that has garnered significant attention is security. Due to the diverse and complex nature of real-world scenarios, even LLMs that have undergone safety alignment remain vulnerable to prompt injection attacks~\cite{greshake2023not,liu2023prompt}. Suo \ea{}~\cite{suo2024signed} proposed Signed-Prompt as a defense against sensitive prompt injection attacks. However, Liu \ea{}~\cite{liu2024demystifying} identified an attack technique that exploits prompts to achieve remote code execution (RCE). Similarly, Evertz \ea{}~\cite{evertz2024whispers} outlined two methods for extracting confidential information through malicious prompts: \textit{(a)} inducing the model to leak data by disguising malicious input as legitimate user queries and \textit{(b)} injecting malicious data into external tools to hijack the model’s behavior, leading it to execute unsafe operations. These examples demonstrate the vulnerability of prompt environments in model integration settings, as observed by Zhan \ea{}~\cite{zhan2024injecagent}. Although various studies have proposed effective defense mechanisms, including benchmarks~\cite{liu2024formalizing,piet2024jatmo,zhan2024injecagent}, fine-tuning~\cite{piet2024jatmo}, prompt filtering~\cite{pedro2024prompt}, and LLM-based defenses~\cite{zhong2025rtbas,piet2024jatmo}, recent studies~\cite{lee2024prompt,panterino2024dynamic,debenedetti2025agentdojo,hui2024pleak,liu2024automatic} highlight their limitations, which underscores the need for further investigation into emerging attack vectors and the development of more robust security strategies to safeguard LLMs in the real world.


In addition to direct attacks targeting the models themselves, the communication protocols facilitating model integration and agent interaction can also become potential entry points for security threats. Protocols such as the \textbf{MCP} (for agent-tool interaction), \textbf{ACP}, \textbf{A2A}, and \textbf{ANP} (for inter-agent communication in various contexts) each introduce specific security risks throughout their lifecycle. For instance, MCP servers, which mediate agent access to external tools, face risks including code injection, backdoor implantation, and installer spoofing, which may lead to information leakage or the incorrect execution of actions by LLMs~\cite{hou2025model}. Radosevich~\ea~\cite{radosevich2025mcp} identified that MCP servers are particularly vulnerable to malicious code execution (MCE), remote access control (RAC), and credential theft (CT). Protocols like A2A, ACP, and ANP, designed for communication between agents, face distinct challenges such as identity spoofing (e.g., Agent Card spoofing in A2A, DID spoofing in ANP), message tampering, unauthorized capability injection, and session hijacking, impacting secure task delegation and coordination~\cite{ehtesham2025survey}. To mitigate these protocol-specific vulnerabilities, research is exploring various defense mechanisms tailored to each protocol's interaction model and lifecycle, including authentication, auditing, secure configuration strategies based on theoretical threat analysis, cryptographic signing of manifests and messages, and robust access control mechanisms~\cite{ehtesham2025survey, narajala2025enterprise, kumar2025mcp}. Although research on the security of these emerging protocols remains limited, existing studies demonstrate that their security issues should not be overlooked, and future work should place greater emphasis on addressing these challenges across the diverse landscape of agent interoperability protocols.

\subsubsection{Challenges}

We categorize the key challenges in model integration into four main aspects: \textbf{prompt transformation}, \textbf{alignment in multimodal systems}, \textbf{multi-model collaboration}, and \textbf{security concerns}. 

\textbf{Prompt transformation.}  
In both multimodal settings and LLM integration with external models and tools, prompts are no longer limited to a single information source. As user input environments become increasingly complex, inputs may include images, audio, text, and so on, which indicates they need to be converted into a common modality (e.g., image-to-audio transformation). Similarly, RAG and knowledge graphs contain vast amounts of structured and unstructured data, requiring the extraction of relevant content for prompt construction. Consequently, transforming diverse information into an input format that LLMs can effectively process is crucial.

\textbf{Multimodal alignment.}  
Achieving effective alignment across different modalities (e.g., text, vision, and audio) remains a fundamental challenge in multimodal models. Failure to accurately transform and synchronize information between modalities can lead to severe performance degradation or hallucination effects. This alignment process requires not only semantic consistency but also the synchronization of internal representation spaces to facilitate cross-modal understanding. While significant research has been conducted on aligning visual and textual information, studies on other data types, such as sensor data and time-series information, remain underexplored.  

\textbf{Multi-model collaboration.}  
The challenges associated with multi-model collaboration can be broadly categorized into two key aspects. First, differences in model capabilities, preferences, and underlying values must be aligned. Otherwise, discrepancies in understanding may lead to reasoning errors during collaboration. Second, coordination challenges arise in workflow management, inter-model communication, and the allocation of computational resources. For instance, assigning simpler tasks to less resource-intensive models while reserving complex tasks for more powerful models can improve inference efficiency and reduce computational costs. However, real-world implementations involve additional complexities, such as quantifying task difficulty, designing model selection strategies, and managing concurrent execution across models. Addressing these challenges requires significant advancements in multi-model coordination and scheduling mechanisms.  

\textbf{Multi-agent system scaling.}
During the scaling of multi-agent systems (MAS), two primary challenges arise: the rapid increase in computational resource demands and the significant growth in complexity of communication and coordination. When the number of agents grows, the system becomes more resource-intensive.  This is because even a single LLM consumes a lot of resources, and adding more agents means extra computational and storage overhead for each one. Additionally, the complexity of communication and collaboration also escalates rapidly with system expansion. Since agents have autonomous decision-making and execution capabilities, ensuring the correctness and consistency of their decisions becomes increasingly complex, exhibiting a nonlinear and potentially exponential growth in complexity. Furthermore, cross-agent communication, complex decision-making, task decomposition, and scheduling become critical and significantly more challenging in large-scale systems. Thus, controlling computational and collaboration costs while scaling remains a fundamental issue that must be addressed in the design of multi-agent systems.

\textbf{Model integration security.}  
A primary security threat in model integration is prompt injection attacks, which can be exploited to extract sensitive information, inject misleading content, or manipulate models into executing unintended actions. While model integration enhances overall system capabilities, it also introduces the potential risks associated with such attacks, underscoring the necessity for robust defense mechanisms. In addition to prompt injection targeting the models, the various interaction protocols used in model integration also introduce new attack surfaces. Protocols standardizing agent-to-external system communication, such as MCP, are susceptible to risks like code injection, remote access control, and credential theft within their server implementations and communication channels, which can result in information leakage or incorrect execution of actions by LLMs. Protocols facilitating agent-to-agent or multi-model communication (e.g., A2A, ACP, ANP) face distinct security challenges related to ensuring message authenticity, authorizing interactions between peers, and preventing malicious coordination. Although research on the security of these emerging protocols is still in development, existing studies highlight the need to develop appropriate authentication, auditing, and configuration mechanisms tailored to each protocol's specific interaction model to mitigate vulnerabilities across the integrated system.

\subsubsection{Road Ahead}\label{sec:model_integration_ra}

Although the challenges mentioned above may appear distinct, they are inherently interconnected due to the nature of model integration, which involves coordinating multiple models or integrating models with external tools. To collectively address these issues, we propose an \textbf{intelligent prompt and secure framework}. This framework consists of three core modules: a \textbf{prompt module}, a \textbf{routing module}, and a \textbf{security module}, each of which presents opportunities for future research and advancements. 

\textbf{Prompt module.}  
This module serves two primary functions: \textbf{prompt generation} and \textbf{prompt filtering}. The prompt generation process must facilitate cross-modal information transformation, extract relevant knowledge from external sources, and ultimately generate optimized input prompts. To achieve this, the module must incorporate system monitoring, contextual understanding, and adaptive decision-making. Furthermore, to ensure effective multimodal alignment, the framework should align diverse data types~\cite{koh2024generating} with LLM representation spaces and potentially leverage token-free Transformer architectures~\cite{pagnoni2024byte} to enhance expressive capacity.  In addition to prompt generation, robust \textbf{prompt filtering} mechanisms are essential for preventing prompt injection attacks. The module must filter potentially malicious or biased prompts before processing, whether they originate from external sources or are generated by the LLM itself. Such defenses are particularly critical in mitigating advanced attack strategies, as identified by Lee \ea{}~\cite{lee2024prompt}, where LLM-generated outputs can inadvertently function as adversarial prompts. 
To achieve this, advanced content-aware techniques are needed to detect and filter malicious or biased content, either based on predefined rules, LLM outputs, or patterns learned by LLMs from data.


\textbf{Routing module.}  
The routing module is responsible for orchestrating LLMs and other subsystems within the framework. It must perform task decomposition, analyze task complexity, generate execution pathways, and select appropriate models or subsystems to execute tasks either in parallel or sequentially. Additionally, it dynamically manages computational resources, enabling efficient scheduling and adaptive reasoning across multi-model, multimodal, and multi-agent systems, thereby enhancing the model integration framework’s overall performance and problem-solving capabilities.  
To achieve these objectives, different protocols and techniques are needed depending on the nature of the interaction. 


For standardizing the interaction between an AI model (agent) and external tools, data sources, or services, often referred to as agent-tool invocation, leveraging the \textbf{MCP} is key. MCP provides a unified interface standard for accessing tools, resources, and prompts, enabling the efficient management of tool and model invocation processes. This standardizes the flow of requests from agents to external capabilities, allowing agents to seamlessly integrate diverse external functionalities by simply adhering to the protocol.
For coordination and collaboration between multiple agents or models, often referred to as agent-to-agent or multi-model communication, specific communication protocols are required to handle message exchange, task delegation, negotiation, and collaborative execution. Protocols like \textbf{A2A}, \textbf{ACP}, and \textbf{ANP} are designed for various forms of inter-agent communication, providing frameworks for performative messaging, capability discovery, and secure peer interaction in different deployment contexts (e.g., within trusted organizational boundaries for A2A, brokered communication for ACP, decentralized open networks for ANP). These protocols are crucial for mitigating semantic inconsistencies and information transmission errors resulting from heterogeneous communication methods among interacting agents or models, thereby enabling the scalable coordination of multi-agent systems.
Beyond communication protocols, achieving robust coordination in multi-agent systems requires sophisticated scheduling and collaboration mechanisms. Techniques from \textbf{swarm intelligence} such as bee algorithm, ant colony optimization (ACO), and particle swarm optimization (PSO), as well as multi-agent game-theoretic models from \textbf{game theory}, can be adopted to optimize task allocation, cooperation strategies, and autonomous decision-making among agents (or models) within large-scale environments. These approaches complement the communication protocols by providing the intelligence layer for complex multi-agent coordination, thus driving LLM-based multi-agent systems toward greater efficiency, scalability, and resource optimization.

\textbf{Security module.}  
Closely interacting with both the prompt and routing modules, the security module acts as a critical protective layer against various threats in the integrated system. It mitigates direct model risks such as prompt injection through techniques including self-supervised anomaly detection, static analysis, sandboxed inference, and reinforcement learning-driven adversarial defense. Furthermore, it supports data isolation and secure interaction mechanisms within complex workflows (including multi-model or multi-agent collaboration), preventing sensitive information leakage and inhibiting the propagation of attacks between components.

Besides, securing the diverse interactions within this framework relies heavily on addressing vulnerabilities in the underlying communication protocols. For \textbf{agent-to-external system interactions} standardized by protocols like MCP, security concerns focus on securing the channel between the agent and the tool server, which involves mitigating risks such as code injection, remote access control, credential theft, and ensuring the integrity of tool invocations and data exchange~\cite{hou2025model, radosevich2025mcp}. To address these concerns, the security module can incorporate measures such as firewall-based server protection, access control, and authentication for tool usage, as well as logging and auditing of tool interactions~\cite{narajala2025enterprise, kumar2025mcp}.
For \textbf{agent-to-agent communication and collaboration} facilitated by protocols such as ACP, A2A, and ANP, security measures are needed to ensure trustworthy interactions between autonomous agents. Challenges here include verifying agent identity, ensuring message authenticity and integrity, authorizing peer-to-peer actions, preventing unauthorized capability injection, and securing the coordination process itself. These protocols incorporate mechanisms like digital signatures, strong authentication (e.g., DIDs, mutual TLS), access control policies for agent skills, and secure session management~\cite{ehtesham2025survey}. The security module can integrate and manage these protocol-specific security features to provide threat detection and defense across all interaction types within the integrated system.

\subsection{Model Compression}\label{sec:model_compression}

\subsubsection{Research Status} 

Model compression aims to reduce the number of parameters or the memory footprint of a model. We categorize it into three main approaches: \textbf{quantization}, \textbf{KD}, and \textbf{pruning}, the same as ~\cite{zhu2024survey,liu2024contemporary,chavan2024faster}.  

\textbf{Quantization} techniques reduce the bit-width of LLM parameters to lower memory usage. Post-training quantization methods typically target weights, activations, or KV cache, using either floating-point or integer formats to reduce runtime memory requirements significantly. For instance, Dettmers \ea{}~\cite{dettmers2022gpt3} compressed both weights and activations to 8-bit integers. More aggressive approaches, such as QuIP~\cite{chee2024quip} and KIVI~\cite{liu2024kivi}, further reduce weights and KV cache to as few as 2 bits. To enhance compression effectiveness, methods such as COMET~\cite{liu2024comet} and QServe~\cite{lin2024qserve} simultaneously quantize weights, activations, and KV cache.  However, aggressive quantization often results in significant performance degradation. To address this issue, researchers have explored retraining quantized models to mitigate performance loss, even if there is a cost of additional training overhead. Techniques such as LLM-QAT~\cite{liu2023llm}, L4Q~\cite{jeon2024l4q}, and QLoRA~\cite{dettmers2024qlora} integrate KD or PEFT to alleviate this overhead. L4Q~\cite{jeon2024l4q} achieves post-training performance comparable to fine-tuned models but is limited to weight quantization, restricting its overall compression potential.  Additionally, Huang \ea{}~\cite{huang2024empirical} reported severe performance degradation in quantized versions of the LLaMA3 model, highlighting the limitations of current quantization techniques in balancing model compactness and performance. These findings underscore the ongoing need for advancements in quantization methodologies to ensure both efficiency and effectiveness.

\textbf{KD} aims to transfer knowledge from a more capable teacher model (TM) to a smaller student model (SM), enabling the latter to achieve comparable performance with reduced computational costs. The challenges in KD can be broadly categorized into two key aspects: \textit{(a)} how to extract knowledge and \textit{(b)} how to learn the extracted knowledge effectively.  A straightforward approach to knowledge extraction is to provide the TM with input data and use its outputs as distilled knowledge~\cite{zhao2024multistage,zhou2024teaching,peng2024metaie}. However, this method heavily depends on the capabilities of the TM~\cite{zhou2024teaching} and is constrained by the diversity and scale of the instruction set~\cite{li2024can}, potentially leading to poor generalization in the SM~\cite{shirgaonkar2024knowledge,shumailov2023curse}. To address these limitations, researchers have integrated data cleaning with synthetic data generation to improve instruction set quality~\cite{yu2024wavecoder,ding2024data,wang2024templatemattersunderstandingrole}. Additionally, self-knowledge techniques, wherein the SM generates new knowledge without relying on the TM, have been explored~\cite{chen2024selfplayfinetuningconvertsweak,yang2024rlcdreinforcementlearningcontrastive}. However, self-knowledge methods are susceptible to inherent biases and hallucinations~\cite{stechly2024selfverificationlimitationslargelanguage,yang2024reinforcingthinkingreasoningenhancedreward}, posing significant challenges to their effectiveness.  Instruction following (IF) is a widely used technique in KD, yet it faces limitations, such as dependency on high-quality datasets~\cite{li2024can} and difficulty in capturing the TM’s reasoning process. To overcome these challenges, techniques like chain-of-thought (CoT) prompting~\cite{wang2024selftrainingdirectpreferenceoptimization,ning2024llmslearnteachingbetter} have been employed to enhance the SM’s reasoning capabilities. Recently, Choi \ea{}~\cite{choi2024embodiedcotdistillationllm} proposed a method that integrates LLM reasoning decomposition and planning capabilities with knowledge graph-augmented reasoning. Their approach, implemented in the lightweight framework DeDer, successfully distills reasoning skills into a compact model, demonstrating the potential of KD to enable resource-constrained devices to perform complex tasks.

\textbf{Pruning} aims to enhance model efficiency and reduce computational overhead by removing redundant neurons or weights. It can be categorized into structured pruning, semi-structured pruning, and unstructured pruning.  Structured pruning methods focus on removing entire structures, such as neurons, channels, or attention heads, to maintain computational efficiency. Ma \ea{}~\cite{ma2023llm} proposed a notable structured pruning approach, LLM-Pruner, which constructs a structural dependency graph of the LLM, groups parameters accordingly, identifies unimportant groups for pruning, and subsequently restores model performance using LoRA. This method requires only 590k samples and three hours of training. However, combining structured pruning with PEFT introduces additional training overhead and can lead to performance degradation~\cite{zhao2023cpet}. To mitigate these issues, Zhao \ea{}~\cite{zhao2024apt} introduced an adaptive pruning strategy, which removes parameters irrelevant to fine-tuning tasks from the pretrained model while incorporating task-specific parameters via distillation, thereby improving LLM performance with reduced computational overhead.  Semi-structured pruning provides a balance between flexibility and efficiency by enforcing structured sparsity patterns at the matrix level. NVIDIA introduced a 2:4 structured sparsity technique, which retains 2 out of every 4 weights in matrix computations, effectively accelerating inference~\cite{mishra2021accelerating}. However, this approach can inadvertently remove critical weights, resulting in a noticeable degradation of accuracy~\cite{tan2024wrp}. To address it, Tan \ea{}~\cite{tan2024wrp} proposed a method that selectively retains essential weights while maintaining the 2:4 sparsity ratio, thereby improving accuracy retention.

\subsubsection{Challenges}

The aforementioned discussion on model compression highlights several key challenges associated with \textbf{quantization}, \textbf{KD}, and \textbf{pruning}.  

For \textbf{quantization challenges}, while existing techniques effectively reduce memory consumption, they often lead to performance degradation, posing a fundamental trade-off between compression efficiency and model accuracy. Balancing these two aspects remains an open challenge. Recent advancements have pushed parameter bit-width to its lower limits, with some approaches reducing it to as few as 2 bits. However, the precise relationship between quantization levels and performance degradation remains unclear. Several studies~\cite{huang2024empirical,huang2024good,gong2024makes} have conducted empirical analyses on the impact of quantization on model performance. However, current investigations remain insufficient, particularly in open-source models such as the LLaMA family. As noted by Jin \ea{}~\cite{jin2024comprehensive} and Yao \ea{}~\cite{yao2024exploring}, existing research primarily focuses on a limited range of models and quantization techniques, leaving significant gaps in understanding the broader implications of extreme quantization.  

Regarding \textbf{KD challenges}, the primary challenges lie in both knowledge extraction and knowledge learning. For knowledge extraction, direct input-based methods offer a straightforward approach; however, they often yield suboptimal training outcomes and weak generalization in SMs. Alternative strategies, such as self-knowledge and instruction following, are hindered by biases, hallucinations, and dataset limitations. In terms of knowledge learning, acquiring abstract capabilities such as reasoning and generalization remains a significant challenge. While CoT prompting has been employed to enhance reasoning abilities in SMs, research on improving other abstract skills remains scarce, highlighting an area for further exploration.  

Finally, \textbf{pruning} presents challenges primarily in parameter selection. The removal of essential parameters can lead to severe performance degradation, yet accurately distinguishing between critical and redundant parameters remains an open research problem. Developing more reliable pruning criteria and adaptive selection mechanisms is crucial to mitigating the risks associated with aggressive parameter reduction.

\subsubsection{Road Ahead}

The challenges mentioned above highlight the key limitations in model compression. While these challenges are primarily studied within the domain of ML, they can also be addressed from an SE perspective. Overall, achieving \textbf{compact and efficient model compression} remains a critical objective for future advancements, necessitating deep optimization of quantization, KD, and pruning, as well as their integration with other optimization techniques to achieve complementary benefits.  

\textbf{Quantization.}  
Existing quantization techniques struggle to simultaneously support weight, KV cache, and activation quantization while maintaining model performance. If all three components can be effectively quantized while leveraging KD or PEFT to recover performance losses, it would enable a better trade-off between computational efficiency and model compactness. Additionally, conducting a comprehensive empirical study on the impact of different quantization levels on model performance could provide valuable insights for future research. Further investigation is needed to assess the effects of quantization across various downstream tasks, quantization techniques, and model sizes. Developing an automated evaluation framework or plugins for quantization impact that is capable of fine-grained performance loss analysis and automated optimization recommendations would be a significant step forward.  

\textbf{KD.}  
Beyond single-model KD, future research should explore the efficient integration of knowledge across multiple models and diverse sources. Given the dynamic and multi-source nature of real-world knowledge, designing a distributed KD training framework that supports multi-node collaboration and asynchronous updates could significantly enhance LLMs' generalization and knowledge update capabilities. Such a framework would be particularly beneficial in federated learning scenarios, enabling efficient KD in decentralized environments. However, despite extensive research in this area~\cite{pang2024federated,qin2024knowledge}, several engineering challenges remain, including issues related to data heterogeneity~\cite{huang2024overcoming,yang2023fedfed,zhu2021data,zhang2022fine}, device heterogeneity~\cite{morafah2024towards,singh2023personalized}, and high communication costs~\cite{gad2024communication,gad2024joint}. Furthermore, to incentivize knowledge sharing among different nodes, blockchain technology could be integrated into the framework, which would not only provide a mechanism for incentivization but also enable knowledge traceability, mitigating the risks associated with malicious nodes injecting harmful information.  

\textbf{Pruning.}  
Semi-structured pruning techniques have demonstrated promising potential, and future research may focus on intelligent weight selection, which requires a deeper understanding of which weights and knowledge are essential for model functionality, which could potentially leverage loss functions, activation functions, or other indicators. Zhang \ea{}~\cite{zhang2024pruning} explored the use of loss functions to distinguish between domain-specific and general knowledge, paving the way for more precise pruning strategies. Additionally, integrating model testing techniques could enhance pruning effectiveness by monitoring performance, which would help prevent severe degradation or the emergence of errors, thereby enabling dynamic pruning adjustments to maintain model reliability while maximizing efficiency.

\subsection{PEFT}\label{sec:peft}
\subsubsection{Research Status} 

PEFT is a technique that updates only a subset of a model’s parameters during fine-tuning, thereby achieving high efficiency while avoiding full-parameter modifications. We categorize PEFT methods  into three main types: \textbf{additive}, \textbf{reparameterized}, and \textbf{selective} approaches, the same as~\cite{han2024parameter,xin2024parameter}.  

\textbf{Additive PEFT} techniques preserve the original model parameters while introducing additional trainable parameters to adapt the model to downstream tasks. Representative methods in this category include adapter layers and soft prompt.  The trained adapter modules are inserted into the model as additional Transformer layers, reducing the number of modified parameters. Based on the insertion strategy, adapters can be classified into sequential adapters and parallel adapters. Sequential adapters primarily focus on adapting to specific tasks, while parallel adapters combine the outputs of both the adapter and the main model, making them better suited for complex scenarios~\cite{kim2024hydra}. However, adapters introduce additional modules, increasing model complexity, maintenance overhead, and inference latency~\cite{ruckle2020adapterdrop}.
In contrast, the soft prompt method appends a set of learnable vectors, aligned with the embedding layer, to the input prompt. These vectors guide the LLM to perform downstream tasks more effectively without modifying the model architecture~\cite{wang2024adapting}. Unlike adapters, soft prompt avoid additional structural complexity and inference overhead and can be transferred between different models and tasks~\cite{xusoft,vykopal2024soft}. However, soft prompt do not fundamentally enhance the model’s capabilities, as they primarily rely on the LLM’s inherent reasoning capabilities~\cite{wang2024universality}. Moreover, soft prompt are vulnerable to adversarial attacks, particularly prompt injection attacks~\cite{nazzal2024promsec,yao2024poisonprompt}. Compared to standard prompt-based attacks, soft prompt manipulations are more likely to bypass a model’s safety alignment mechanisms and induce unintended behaviors~\cite{schwinn2024soft,yang2024sos}. For example, the malicious soft prompt can lead to unintended data leakage, including the inadvertent exposure of sensitive information from the training corpus~\cite{kim2024propile}.  

\textbf{Selective PEFT} fine-tunes a model by masking a portion of its parameters, similar to pruning. It can be broadly categorized into structured masking and unstructured masking, both of which aim to identify an optimal subset of parameters for fine-tuning. Representative approaches include gradient-based methods~\cite{chekalina2024sparsegrad,li2024gradient,yang2024s}, data-driven methods~\cite{dong2024data,Das2023UnifiedLS}, and search-based methods~\cite{chang2024bipeft,zhou2024autopeft}.  The primary advantage of selective PEFT is that it does not increase the inference cost of the LLM, making it an attractive alternative to other PEFT techniques. However, the complexity of parameter selection strategies introduces significant challenges in model development and debugging. Additionally, Ploner \ea{}~\cite{ploner2024parameter} observed that randomly selected parameter subsets such as those employed in LoRA often yield performance comparable to carefully designed selection strategies. Their findings raise questions regarding the practical benefits of extensive debugging efforts, given the marginal improvements achieved over random parameter selection.

\textbf{Reparameterized PEFT} modifies the model's parameterization to enable more efficient adaptation. Among these methods, LoRA is a widely adopted approach. LoRA employs low-rank decomposition to train a separate module, which is then used to reparameterize specific model weights. A single model can incorporate multiple LoRA-trained modules, allowing for the selection of different module combinations during inference based on specific requirements. Despite its efficiency, LoRA presents two primary challenges:  improving LoRA’s performance and selecting appropriate LoRA modules.  \textit{(a) Performance optimization.}  
Numerous techniques have been proposed to enhance LoRA’s effectiveness, including dynamic rank adjustment~\cite{valipour2022dylora,zhang2023adalora}, earning rate optimization~\cite{hayou2024lora+}, and regularization strategies to mitigate overfitting~\cite{lin2024lora,wang2024lora}. However, the extent of these improvements remains constrained. Zhang \ea{}~\cite{zhang2024scaling} found that LoRA’s performance is predominantly influenced by the inherent capabilities of the base model, suggesting that optimizations at the LoRA level offer only limited benefits.  \textit{(b) Module selection and inference efficiency.}  
Optimizing LoRA module selection can significantly reduce LLM inference latency and enhance overall performance. For instance, Kong \ea{}~\cite{kong2024lora} observed that the insertion of LoRA modules leads to fragmented CUDA kernel calls, severely degrading inference efficiency. To address this, they proposed a novel token-wise routing strategy to minimize unnecessary kernel invocations. Similarly, Wu \ea{}~\cite{wu2024dlora} introduced a dynamic switching mechanism between merged and unmerged models to reduce inference latency in model as a service (MaaS) scenarios. Their approach further integrates batching techniques and a request-adapter co-migration strategy to improve GPU resource utilization and overall service performance. \textit{(c) security considerations.}  
Despite its advantages, LoRA introduces security concerns due to the additional fine-tuning it requires. Liu \ea{}~\cite{liu2024lora} demonstrated that open-source LoRA adapters are vulnerable to backdoor attacks. Moreover, Xu \ea{}~\cite{hsu2024safe} highlighted that even when training datasets do not contain malicious data, aligned LLMs remain susceptible to adversarial threats. To mitigate these risks, they proposed Safe LoRA, which constrains LoRA updates using a projection operation, ensuring that parameter updates align with a predefined security-preserving matrix, thereby enhancing robustness against adversarial manipulations.

\subsubsection{Challenges}

The research mentioned above highlights several key challenges associated with PEFT. We categorize these challenges as follows:  

\textbf{Additive PEFT challenges.}  
While adapter layers facilitate seamless integration with LLMs, they also introduce additional complexity in system maintenance. The selection, combination, and interconnection of different adapter layers pose significant engineering challenges, yet they also present opportunities for further advancements. Moreover, the insertion of adapter layers inevitably increases inference latency, which can be detrimental to latency-sensitive applications. In contrast, soft prompt mitigate these performance concerns but raise security and privacy risks, as they may inadvertently expose sensitive or private data from the training process.  

\textbf{Selective PEFT challenges.}  
The primary challenge in selective PEFT lies in selecting the appropriate parameters. Simple selection strategies may fail to achieve optimal fine-tuning results, whereas more sophisticated selection mechanisms not only introduce additional development and debugging overhead but also do not necessarily outperform random selection, which raises concerns regarding the feasibility and practical benefits of selective PEFT.  

\textbf{LoRA challenges.}  
As a representative reparameterized PEFT method, LoRA enables efficient fine-tuning but faces limitations related to inference efficiency and security vulnerabilities. The integration of LoRA could lead to fragmented CUDA kernel calls, thereby reducing inference efficiency. Additionally, LoRA-trained adapters have been demonstrated to be susceptible to backdoor attacks. Although ongoing research aims to enhance LoRA’s robustness and efficiency, substantial challenges remain, limiting its applicability in security-critical and latency-sensitive scenarios.

\subsubsection{Road Ahead}

To address the challenges mentioned above, there are several research directions for further exploration. For challenges associated with adapter layer insertion, an \textbf{adaptive adapter architecture} could be developed to facilitate adapter selection, composition, and integration. Hu \ea{}~\cite{hu2023llm} proposed a broad adapter integration framework. However, their approach lacks considerations for composition design and optimization at deployment. Future research could focus on automated optimization, modularization, and standardization of adaptive adapter architectures. For instance, standardized APIs for adapter layers, dynamic adapter loading mechanisms that activate specific adapters only when necessary, and optimized caching strategies could significantly enhance inference efficiency.
Additionally, automatic search techniques could be employed to determine the optimal adapter combinations, adapter depth, and activation strategies. While such techniques have been extensively studied in vision models~\cite{zhu2024task,yu2023lape,yin2024parameter}, their application in LLMs remains underexplored. Notably, research on adaptive architectures is also applicable to LoRA, since it can be seen as a special form of adapter that modifies the original Transformer's weights instead of introducing additional layers. For example, Wu \ea{}~\cite{wu2024dlora} proposed a method for dynamically merging and migrating LoRA adapters to enhance the throughput of LLM servers.  

For challenges related to parameter selection in selective PEFT, similar to pruning, \textbf{parameter testing and recommendation tools} could be leveraged to evaluate the impact of different hyperparameter configurations on model performance, which would enable the development of more effective parameter selection strategies while also facilitating the identification of optimal hyperparameter combinations.  

PEFT security presents two key challenges: \textit{(a)} soft prompt may expose private data, and \textit{(b)} adapters (including LoRA adapters) are vulnerable to backdoor attacks.  To mitigate risks associated with soft prompt, \textbf{soft prompt filtering techniques}, as discussed in \S\ref{sec:model_integration_ra}, could be extended to prevent inadvertent data leakage. These techniques must ensure that filtered prompts retain both security and semantic fidelity. However, unlike standard prompt-based attacks, soft prompt manipulate the embedding layer, making them more challenging to defend against using traditional LLM security mechanisms~\cite{schwinn2024soft}. Therefore, further research on protective measures for embedding layers is necessary.  Regarding backdoor vulnerabilities in adapters, a potential solution could involve \textbf{data filtering and detection} inspired by existing research on mitigating data poisoning in open-source models. For instance, verifying whether LoRA’s training data has been compromised could leverage data provenance techniques, such as traceability analysis and trusted data sources, as outlined in \S\ref{sec:data_security_ra}. Additionally, adversarial training could be employed to enhance the robustness of LoRA-based adaptations, thereby improving security and reliability.

\section{Testing and Evaluation}\label{sec:testing_evaluation}

\begin{figure}[htbp]  
    \centering     
    \includegraphics[width = \textwidth]{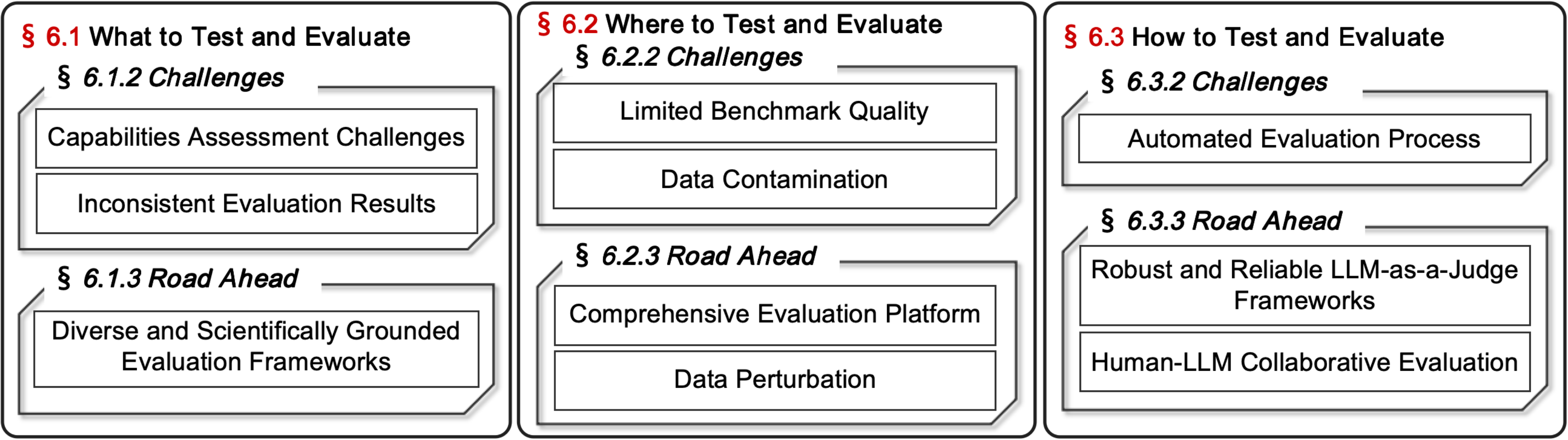}  
    \caption{Challenges and Road Ahead in \S\ref{sec:testing_evaluation} Testing and Evaluation.}
    \label{fig:testing_road_ahead}

\end{figure}

The testing and evaluation of LLMs pose multifaceted challenges, as shown in \autoref{fig:testing_road_ahead}. Inspired by Chang \ea{}~\cite{chang2024survey}, we propose that these challenges can be systematically analyzed through three critical dimensions: \textbf{what}, \textbf{where}, and \textbf{how} to test and evaluate LLMs. These dimensions extend beyond the technical assessment of model performance to encompass broader considerations related to model deployment and real-world usage. The complex relationship between these factors highlights the need for systematic methodologies and innovative evaluation frameworks to ensure the reliability, robustness, and fairness of LLMs.

\subsection{What to Test and Evaluate}
\subsubsection{Research Status}

The evaluation of LLMs spans multiple dimensions. From an application-driven perspective, researchers have explored LLM performance across diverse domains, including medicine~\cite{cai2024medbench,dakhel2024effective}, education~\cite{yan2024practical}, SE~\cite{liu2024your}, and finance~\cite{xie2024pixiu}. In domains requiring advanced cognitive capabilities or creativity, studies have assessed LLMs’ effectiveness in scientific research~\cite{sun2024scieval} and creative tasks~\cite{chakrabarty2024art}. Additionally, evaluations have focused on intrinsic model attributes, such as bias~\cite{rottger2024political}, reasoning capabilities~\cite{wang2024causalbench}, and planning capabilities~\cite{momennejad2024evaluating}.  Despite these efforts, there are still two significant challenges: 

\textbf{Difficulty in assessing certain capabilities.}  
Many LLM capabilities are inherently difficult to quantify. Generative tasks, such as dialogue generation and writing, exhibit a high degree of subjectivity, making them challenging to evaluate using traditional objective metrics (e.g., accuracy). Scientifically quantifying abstract factors such as “creativity,” “relevance,” or “user satisfaction” remains an open research problem~\cite{kenthapadi2024grounding}. Additionally, certain attributes, such as reasoning, are difficult to observe directly. It is often unclear whether an LLM derives its responses through actual reasoning processes or merely retrieves relevant information from its knowledge base~\cite{valmeekam2024planbench,amirizaniani2024can}.  

\textbf{Inconsistency in evaluation results.}  
Variations in evaluation methodologies and metrics across different domains and tasks frequently result in inconsistencies in model performance assessments. For example, Gandhi \ea{}~\cite{gandhi2024understanding} identified discrepancies in LLM reasoning test results, highlighting underlying limitations in existing evaluation frameworks. Moreover, Greenblatt \ea{}~\cite{greenblatt2024alignment} documented an issue where models adhere to training objectives during fine-tuning but fail to maintain this alignment in different scenarios, a phenomenon referred to as \textit{alignment faking}. These inconsistencies raise concerns regarding the validity and robustness of current evaluation methodologies~\cite{gandhi2024understanding}, underscoring the need for more comprehensive and rigorous evaluation.

\subsubsection{Challenges}

\textbf{Capabilities assessment challenges}. 
While certain LLM capabilities, such as code completion and mathematical computation, can be evaluated using manually designed test cases, more abstract skills such as reasoning, writing, and planning pose significant challenges for traditional evaluation methodologies. This difficulty arises from the inherent complexity of designing effective test cases, as well as the absence of well-defined quantitative metrics for objectively measuring these higher-order cognitive capabilities.  

\textbf{Inconsistent evaluation results.}  
Another challenge is that assessments of whether an LLM possesses a particular capability or adheres to alignment expectations often yield inconsistent results. One possible explanation is the absence of scientifically rigorous evaluation methodologies, which can lead to discrepancies in judgment across different evaluation frameworks. Another factor is the phenomenon of \textit{alignment faking}~\cite{greenblatt2024alignment}, where models appear to comply with expected behaviors during evaluation but deviate from them in different scenarios, raising concerns about the reliability of existing evaluation techniques.

\subsubsection{Road Ahead}
The challenges mentioned above highlight the limitations of current evaluation methodologies. Future research could focus on developing \textbf{diverse and scientifically grounded evaluation frameworks} that incorporate cross-domain methodologies to ensure comprehensive, reliable, and objective assessments of LLM performance across varied capabilities and tasks.

\textbf{Cross-domain methodologies for capability assessment.}  
A promising direction for improving the evaluation of LLM capabilities is the integration of cross-domain methodologies, which can enhance the scientific rigor of assessment techniques. 
This cross-domain approach holds significant potential. For example, in terms of KD, insights from educational science could be leveraged to assess a TM’s effectiveness in knowledge transfer or an SM’s capability to acquire and generalize learned information. Similarly, logic-based analysis could provide a more rigorous framework for evaluating reasoning capabilities. By adopting scientifically rigorous evaluation methods, researchers can not only improve the assessment of abstract capabilities but also mitigate inconsistencies in evaluations.  

\textbf{Enhancing alignment evaluation through different testing environments.}  
The emergence of \textit{alignment faking}~\cite{greenblatt2024alignment} underscores the influence of evaluation environment inconsistencies, wherein models demonstrate different behaviors under varying conditions. This phenomenon suggests that future evaluation methodologies should account for the impact of varying environments on evaluation outcomes, ensuring consistency and reliability across real-world usage scenarios. Developing adaptive evaluation frameworks that dynamically adjust test scenarios based on user inputs and interaction patterns could improve the robustness of alignment assessments. Additionally, integrating longitudinal evaluation strategies, where models are assessed over extended periods rather than in isolated test cases, could provide deeper insights into the stability of alignment and behavioral consistency. By grounding evaluations in different interaction environments, researchers can ensure more reliable assessments of LLM alignment and generalization capabilities.

\subsection{Where to Test and Evaluate}
\subsubsection{Research Status}

Numerous benchmarks have been developed to evaluate the capabilities of LLMs, each exhibiting distinct characteristics. Many of these benchmarks focus on assessing only a subset of LLM capabilities. For instance, code-related benchmarks such as xCodeEval~\cite{khan2023xcodeeval}, CoderUJB~\cite{zeng2024coderujb}, and CrossCodeEval~\cite{ding2024crosscodeeval} primarily evaluate code understanding, generation, translation, and retrieval. Similarly, reasoning-oriented benchmarks like PlanBench~\cite{valmeekam2024planbench} and LegalBench~\cite{guha2024legalbench} target reasoning capabilities, while multimodal benchmarks such as SEED-Bench~\cite{li2024seed} and MM-SafetyBench~\cite{liu2025mm} assess the performance of multimodal LLMs. Even within these specialized domains, benchmarks often focus on narrower subtasks. For example, within code-related evaluations, benchmarks like HumanEval~\cite{chen2021evaluating}, ClassEval~\cite{du2023classeval}, and EvoCodeBench~\cite{li2024evocodebench} assess Python code generation, whereas JavaBench~\cite{cao2024javabench} and SWE-Bench-Java~\cite{zan2024swe} evaluate Java code generation and repair, respectively.  

Despite their utility, existing benchmarks suffer from several limitations. First, limited test case coverage restricts their comprehensiveness, as many benchmarks contain only a small number of test cases, reducing their capability to provide a holistic assessment of LLM capabilities. Additionally, the lack of standardized benchmark construction guidelines results in inconsistent dataset quality, leading to fragmented evaluation methodologies. As noted by McIntosh \ea{}~\cite{mcintosh2024inadequacies}, many existing benchmarks fail to capture nuanced aspects such as bias, genuine reasoning, and adherence to cultural and ideological norms. 

Another significant challenge is data contamination, where test cases from benchmarks may have been seen by LLMs during training, leading to unrealistically high evaluation scores and overestimated model capabilities. This issue arises due to overlaps between real-world data used for benchmark construction and LLM training datasets. Even manually curated benchmarks such as HumanEval have been found to contain instances that newer models have encountered during training~\cite{yang2023rethinking,riddell2024quantifying}.To mitigate this issue, some studies have proposed ensuring that benchmark data is sampled after the release of LLMs~\cite{feng2024complexcodeeval,khan2023xcodeeval}. However, these efforts have proven insufficient, as contamination can still occur when newer models are trained on datasets that include older benchmark samples~\cite{yang2023rethinking, cao2024concerned}. Data augmentation has gained attention as a potential solution, as it can generate novel data instances that were not present in the original dataset. Zhu \ea{}~\cite{zhu2024dynamic} introduced a psychometric-inspired data augmentation method to evaluate LLMs from multiple perspectives by modifying existing datasets. Additionally, researchers have leveraged LLM-based methods~\cite{xia2024top} to generate augmented datasets, taking advantage of LLMs' advanced language understanding and generation capabilities.

Ultimately, the lack of standardized benchmark usage guidelines leads to significant variability in evaluation quality and scope, resulting in divergent and sometimes inconsistent assessment outcomes. To address this, it is essential to develop a standardized evaluation framework that objectively assesses the reliability, scope, and validity of benchmark-based evaluations. In general, these limitations highlight the need for comprehensive research to develop more robust benchmarks, establish effective anti-contamination mechanisms, and create standardized assessment methodologies to enhance the reliability and fairness of LLM evaluations.

\subsubsection{Challenges}

\textbf{Limited benchmark quality.}  
While the number of benchmarks for evaluating LLMs has grown rapidly, many of them still face fundamental quality limitations. First, their scope is often restricted, even within a specific capability domain; existing benchmarks frequently fail to provide comprehensive coverage. Second, there is a notable difficulty distribution imbalance in test cases—some benchmarks are excessively challenging, while others are too simplistic, making it difficult to accurately assess an LLM’s actual capabilities. Third, benchmarks rapidly become outdated, as LLMs continue to advance, many existing benchmarks lose their effectiveness over time, necessitating continuous updates and refinements to remain relevant.  

\textbf{Data contamination.}  
As discussed before, data contamination can significantly distort evaluation results by introducing test cases that LLMs may have seen during training. Although mitigation strategies, such as post-release sampling and sample rephrasing, have been proposed, these approaches remain imperfect, highlighting the need for more robust methodologies that can effectively minimize data contamination while ensuring the validity and reliability of assessments.

\subsubsection{Road Ahead}

As LLMs continue to evolve, the limitations of existing evaluation benchmarks will persist, posing ongoing challenges. However, the development of a \textbf{comprehensive evaluation platform} could significantly mitigate these issues. Such a platform would provide a dedicated infrastructure for benchmark maintenance, facilitating continuous updates and the integration of new evaluation datasets. By enhancing benchmark reliability and diversity, this approach could help address the inherent shortcomings of current evaluation methodologies. While platforms such as Hugging Face~\cite{huggingface_datasets} offer shared evaluation datasets, they lack effective dataset management, systematic benchmark updates, and continuous integration capabilities. Consequently, evaluation dataset quality remains inconsistent, limiting their ability to ensure benchmark reliability. To overcome these challenges, future research should focus on designing an adaptive benchmark management system that enables automated dataset curation, real-time benchmark refinement, and the dynamic integration of newly proposed evaluation metrics.  

Additionally, since many LLM training datasets are not publicly available, the risk of data contamination remains a critical concern. To address this, \textbf{data perturbation} techniques could provide a potential solution. By transforming existing benchmark samples into novel representations while preserving their original semantic meaning, these techniques could generate test cases that are distinct from those encountered during model training. This approach would help reduce the likelihood of evaluation biases, regardless of whether the original benchmark data is sourced from real-world datasets or synthetically generated.

\subsection{How to Test and Evaluate}
\subsubsection{Research Status}

In the evaluation of LLMs, certain capabilities, such as fill-in-the-middle (FIM) performance in the coding domain~\cite{ding2023crosscodeeval,gong2024evaluation,liu2023repobench,wu2024repomastereval}, can be assessed through the automated execution of test cases, yielding objective pass rate metrics. However, more abstract capabilities, such as creativity and reasoning, are significantly more challenging to evaluate automatically and often require human judgment~\cite{chakrabarty2024art,wang2023evaluating,singhal2022large}.  This reliance on manual evaluation introduces two primary issues. First, human judgment is subjective, leading to inconsistencies and reduced reliability in evaluation outcomes. Second, manual assessment is both time-consuming and labor-intensive, making large-scale evaluations impractical and limiting the comprehensiveness of assessments. These challenges highlight the need for developing more scalable and objective evaluation methodologies.

\subsubsection{Challenges}
\textbf{Automated evaluation process.} 
While certain LLM capabilities can be assessed using automated tools, more abstract capabilities such as creativity and reasoning remain challenging to evaluate fully automatically. As a result, manual assessment is often required, introducing subjectivity that reduces the accuracy and reliability of evaluation outcomes. Moreover, the reliance on human judgment limits the feasibility of conducting large-scale assessments, posing a significant barrier to comprehensive and scalable LLM evaluation.

\subsubsection{Road Ahead}

Given the powerful capabilities of LLMs, LLM-based evaluation methods, commonly referred to as LLM-as-a-Judge, have gained increasing attention~\cite{lin2023llm,chu2024automatic,hashemi2024llm,wang2025can,li2024llms}. However, these methods are inherently influenced by the biases present in LLMs~\cite{weirocketeval,kumar2024decoding,ye2024justice}, as well as their intrinsic limitations~\cite{van2024field,weirocketeval}, which may lead to inaccurate evaluation outcomes. Therefore, a key research direction is the development of more \textbf{robust and reliable LLM-as-a-Judge frameworks}, which includes strategies to mitigate biases in evaluation and the integration of multi-model and multi-modal approaches to enhance fairness and reliability. By leveraging diverse models and modalities, these methods could reduce individual model biases and improve overall assessment accuracy. Another promising direction is \textbf{human-LLM collaborative evaluation}, which serves as a compromise between full automation and evaluation accuracy. The goal of this approach is to balance the efficiency of automated evaluation with the careful judgment of human evaluators, thereby producing LLM assessments that are both reliable and interpretable.

\section{Deployment and Operations}\label{sec:model_deployment}

\begin{figure}[htbp]  
    \centering     
    \includegraphics[width = \textwidth]{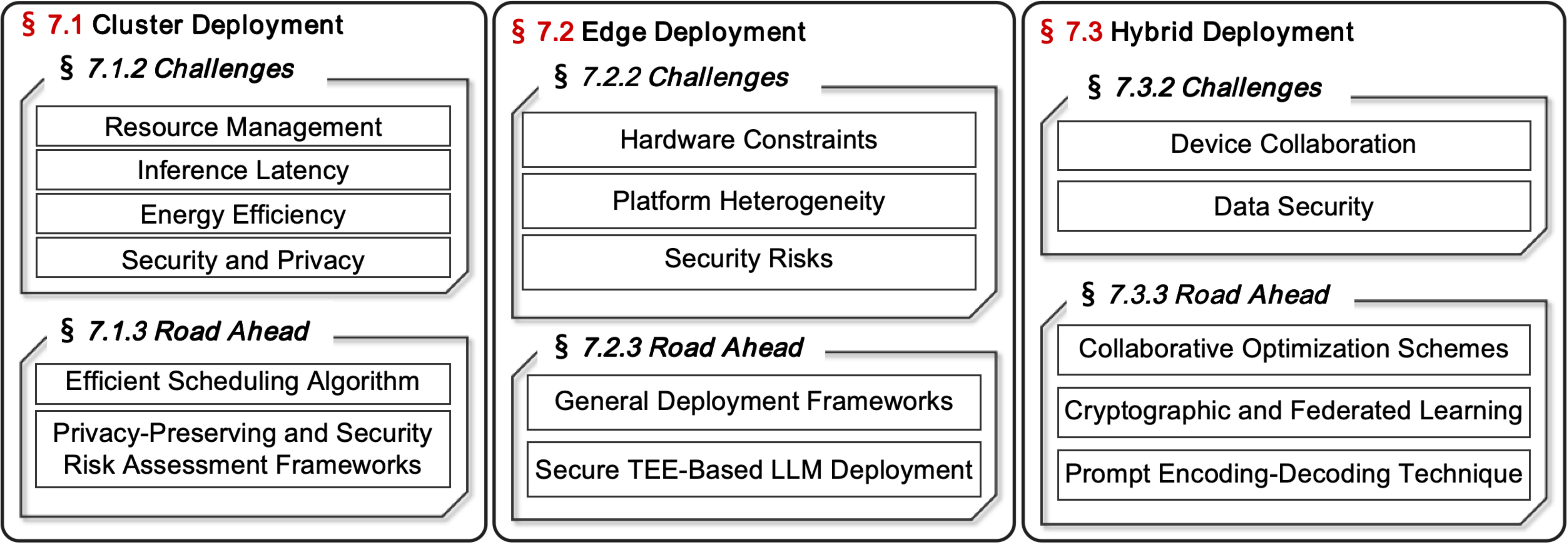}  
    \caption{Challenges and Road Ahead in \S\ref{sec:model_deployment} Deployment and Operations.}
    \label{fig:deployment_road_ahead}

\end{figure}

As discussed in \S\ref{sec:model integration}, the deployment of LLMs presents several challenges, including computational resource constraints, deployment architecture design, and security and privacy concerns. To systematically explore these challenges, we classify LLM deployment into three categories based on the location of computational resources: \textbf{cluster deployment}, \textbf{edge deployment}, and \textbf{hybrid deployment}, which is shown in \autoref{fig:deployment_road_ahead}.

\subsection{Cluster Deployment}

\subsubsection{Research Status}\label{sec:deployment_rs}

Cluster deployment refers to deploying models in high-performance computing clusters, such as the cloud, to leverage distributed computing for large-scale inference. While this approach enables efficient processing of large-scale requests, it also introduces significant technical and operational challenges.

\textbf{Resource management.}  
Managing computational resources in a distributed environment is inherently complex, particularly due to the need for efficient parallelization across heterogeneous hardware (e.g., GPUs, TPUs). To keep systems responsive while evenly distributing tasks, there’s a need for smart scheduling methods and flexible resource scaling~\cite{zhao2024llm}. Hisaharo \ea{}~\cite{hisaharo2024optimizing} redesigned the cluster computing architecture of GPT-Neo and optimized software implementations to enhance inference efficiency. Similarly, Zhao \ea{}~\cite{zhao2024llm} proposed an adaptive algorithm for optimizing LLM inference in heterogeneous environments by dynamically adjusting mixed-precision quantization and GPU allocation strategies to improve throughput. Additionally, LLM-Pilot~\cite{lazuka2024llm} introduced a predictive model that recommends cost-effective hardware configurations,
further improving resource utilization.

\textbf{Inference latency.}  
Despite the high throughput of cluster-based deployment, ensuring consistently low-latency responses under high-concurrency conditions (e.g., thousands of simultaneous requests) remains a significant challenge. Optimized batching strategies, model partitioning techniques (e.g., pipeline heterogeneity~\cite{he2024fastdecode}), and hardware-aware kernel fusion are commonly employed to mitigate computational bottlenecks. Local checkpoint storage~\cite{fu2024serverlessllm} and splitwise techniques~\cite{patel2024splitwise} have been proposed to enhance inference efficiency. However, while checkpoint storage reduces redundant computation, it increases storage overhead, whereas inter-stage computation transfers in splitwise techniques offer only limited latency reductions. Sarathi-Serve~\cite{agrawal2024taming} introduced chunked-prefill and stall-free scheduling techniques, significantly reducing inference latency under high-throughput conditions, with greater optimizations observed for larger models, suggesting their scalability benefits. Zhang \ea{}~\cite{zhang2024edgeshard} explored collaborative edge computing to partition clusters and employed dynamic programming to minimize latency and maximize throughput. However, cross-partition data transfers introduce potential privacy risks, which will be discussed later.

\textbf{Energy efficiency.}  
The substantial energy consumption associated with training and inference on GPU/TPU clusters necessitates effective hardware utilization and memory optimization. To address this issue, Wilkins \ea{}~\cite{wilkins2024hybrid} proposed a technique that dynamically allocates computational resources based on token input-output ratios, thereby reducing energy consumption. 
Hisaharo \ea{}~\cite{hisaharo2024optimizing} and Stojkovic \ea{}~\cite{stojkovic2024dynamollm} introduced optimization algorithms for dynamically managing inference resources to lower energy consumption. However, the search space of such algorithms remains large, prompting Maliakel \ea{}~\cite{maliakel2025investigating} to explore key parameters affecting energy efficiency across different LLMs and tasks. Their findings highlight the need for task-specific hardware optimizations to enhance efficiency further. Additionally, Hewage \ea{}~\cite{hewage2025aging} identified CPU aging as a contributing factor to increased energy consumption and proposed optimization strategies aimed at extending CPU longevity to reduce energy usage.

\textbf{Security risks.}  
Cluster-based LLM deployment is susceptible to security threats, including adversarial attacks on exposed APIs and data leakage in multi-tenant environments~\cite{dash2024zero}. Yang \ea{}~\cite{yang2024first} found that LLM service providers often optimize inference efficiency by sharing KV caches, inadvertently exposing user privacy. Furthermore, Soleimani \ea{}~\cite{soleimani2025wiretapping} demonstrated vulnerabilities in speculative decoding optimizations, introducing a novel side-channel attack capable of extracting LLM token information from encrypted transmissions, such as token size.  To mitigate these risks, encryption techniques such as multi-party computation (MPC)~\cite{rathee2024mpc}, homomorphic encryption (HE)~\cite{zimerman2024power}, and TEE~\cite{mohan2024securing} have been applied to LLM inference. However, while these methods enhance privacy protection, they introduce significant computational overhead, resulting in increased inference latency, which limits their practicality in real-world applications. Additionally, handling non-linear computations such as the \texttt{softmax} function remains a significant challenge in encrypted LLM inference~\cite{zimerman2024power}. 

In addition, weak authentication mechanisms and the imperfect of LLM deployment frameworks introduce further security concerns. Pesati \ea{}\cite{pesati2024security} found that unreliable authentication mechanisms can lead to LLMs being hijacked or manipulated, posing serious security risks. Hou \ea{}\cite{hou2025unveiling} analyzed several popular LLM deployment frameworks such as ComfyUI~\cite{comfyui} and Ollama~\cite{ollama}, and identified widespread model information disclosure vulnerabilities. These issues can also lead to unauthorized access, exploitation of system vulnerabilities, and other security threats. 


\subsubsection{Challenges}

As discussed in \S\ref{sec:deployment_rs}, the challenges of LLM cluster deployment can be categorized into the following four aspects. 

\textbf{Resource management.}  
Efficient resource management is crucial in heterogeneous and distributed environments, requiring sophisticated scheduling algorithms and dynamic scaling mechanisms to balance concurrency and inference latency. The challenge of handling GPUs, TPUs, and other specialized hardware makes performance tuning more difficult, requiring flexible resource allocation methods that adapt to different system demands.

\textbf{Inference latency.}  
Minimizing inference latency is essential not only for improving user experience but also for reducing operational costs. However, optimizing LLM inference presents a systemic engineering challenge that involves architectural design, workload distribution, and hardware utilization. Techniques such as model partitioning, batching strategies, and pipeline parallelism have led to noticeable improvements in performance. However, they still face limitations in scalability and hardware efficiency, leaving many areas open for further refinement.

\textbf{Energy efficiency.}  
LLM inference demands substantial computational resources, resulting in high energy consumption. Existing energy-saving strategies include optimizing resource allocation, extending the lifespan of hardware, and implementing efficient scheduling policies. However, given the vast optimization space, determining the optimal configuration remains a significant challenge.

\textbf{Security and privacy.} 
Cluster-based LLM deployment introduces security and privacy risks, including potential data leakage due to shared KV caches and API calls. Issues such as weak authentication, information disclosure, and unauthorized access further highlight the diversity and complexity of LLM security challenges. Addressing these issues cannot rely on patching individual vulnerabilities or adopting traditional security technologies such as encryption. Instead, it requires a comprehensive security strategy that combines robust access control, fine-grained API governance, secure configuration management, and continuous monitoring. Moreover, securing LLM deployments should be treated as a system-level problem, involving coordination across model architecture, serving infrastructure, and user interaction layers to ensure a defense-in-depth approach.


\subsubsection{Road Ahead} 

We categorize the challenges mentioned above into two key areas: \textbf{LLM cluster inference optimization} and \textbf{LLM cluster inference privacy and security concerns}. 

For LLM inference optimization, existing approaches to addressing high concurrency, low latency, and energy efficiency primarily focus on scheduling design and architectural optimization, making these areas critical for further research. However, the vast number of optimization factors results in a huge search space, complicating the identification of an optimal configuration. A crucial future direction is the development of an \textbf{efficient scheduling algorithm} capable of dynamically exploring this search space and autonomously determining resource allocation and real-time scheduling strategies. Such an algorithm should be designed to optimize multiple objectives simultaneously, ensuring high concurrency, low latency, and reduced energy consumption while adapting to varying workloads and hardware configurations.


Concerning privacy and security concerns, establishing a comprehensive \textbf{privacy-preserving and security risk assessment framework} will be essential in the future. Regarding privacy preservation, differential privacy techniques~\cite{zhang2025no,tong2023inferdpt} show promise by introducing controlled noise into prompts or token representations, obscuring sensitive information. Such techniques offer configurable privacy settings that balance data confidentiality with model utility, thus enhancing the practicality of secure LLM inference. Regarding security risk assessment, it is equally essential to incorporate mechanisms for identifying, assessing, and mitigating security threats. This includes developing unified identity and access management frameworks to enforce fine-grained, role-based permissions, ensuring minimal privilege assignment, and providing standardized authentication interfaces to facilitate integration with third-party services. Additionally, specialized security analysis tools, such as static configuration analyzers, automated endpoint scanning, and risk assessment utilities, should be introduced to systematically address issues related to configuration leakage and interface exposure. These integrated efforts collectively reinforce the robustness of LLM deployment frameworks.


\subsection{Edge Deployment}\label{sec:edge_deployment}

\subsubsection{Research Status}

Edge deployment refers to deploying LLMs on edge devices near the data source, such as smartphones, IoT devices, and edge servers, referred to as on-device LLMs, rather than relying on centralized cloud infrastructure. This approach alleviates the computational burden on cloud-based LLM services while enhancing the quality of commercial LLM applications. However, it introduces challenges, primarily due to memory and computational resource constraints~\cite{dhar2024empirical}.  

\textbf{Hardware constraints.}  
Edge devices typically have limited computing resources (e.g., mobile CPUs and GPUs), restricted memory capacity, and energy constraints, necessitating the use of model compression techniques. However, these techniques often come at the cost of reduced accuracy or robustness. To address this issue, Ma \ea{}~\cite{ma2024era} proposed a quantization technique where each parameter takes values from \{-1,0,1\}, achieving significant improvements in memory efficiency, inference latency, energy consumption, and throughput. Similarly, Li \ea{}~\cite{li2024transformer} introduced four techniques, providing diverse strategies for improving LLM deployment on mobile devices.  

\textbf{Platform heterogeneity.}  
The deployment of LLMs on edge devices is further complicated by platform heterogeneity, as models must be compatible with diverse architectures (e.g., ARM-based chips, NPUs). Several existing solutions facilitate cross-platform deployment. Llama.cpp~\cite{llama_cpp} is a C++ library that supports LLM deployment across various hardware platforms, integrating integer quantization and GPU acceleration. MNN~\cite{jiang2020mnn}, a mobile neural network framework, enables deployment across different backends, with its extension MNN-LLM specifically designed for LLM deployment on mobile devices, PCs, and embedded systems. ExecuTorch~\cite{executorch} is an end-to-end edge inference framework tailored for deploying PyTorch models on edge devices.  

\textbf{Security risks.}  
Edge deployment also introduces heightened security risks, as models are exposed in a white-box manner, making them more susceptible to physical tampering, adversarial inputs, and model stealing~\cite{li2024translinkguard}. However, processing sensitive user information locally reduces the risk of data leakage compared to cloud-based deployment~\cite{zhang2024edgeshard,li2024translinkguard}, highlighting a trade-off between security threats and privacy benefits.

\subsubsection{Challenges}

Edge deployment presents three primary challenges.
\textbf{Hardware Constraints.}  
Unlike centralized servers, edge devices have limited computational capabilities, necessitating model compression before deployment. However, this process involves a trade-off between model size and performance, as aggressive compression techniques can degrade accuracy and robustness.
\textbf{Platform Heterogeneity.}  
The diversity of operating systems and hardware architectures across edge devices complicates deployment, requiring additional driver support and optimization for compatibility. While certain open-source tools facilitate cross-platform deployment, most solutions are tailored to specific LLM ecosystems or hardware platforms, limiting their general applicability.
\textbf{Security Risks.}  
Models deployed on edge devices are effectively exposed in a white-box manner, making them vulnerable to threats such as model extraction and adversarial attacks. Ensuring robust security mechanisms while maintaining efficient inference remains a challenge.

\subsubsection{Road Ahead}

Both \textbf{hardware constraints} and \textbf{platform heterogeneity} are deployment-related challenges that can be addressed through SE solutions. A key direction for future research is the development of \textbf{generalized deployment frameworks}, such as Llama.cpp and MNN, which facilitate seamless LLM deployment across diverse edge devices. These frameworks should not only support cross-platform compatibility but also integrate common model compression and fine-tuning techniques, allowing users to optimize models according to specific deployment requirements. Furthermore, to enhance their applicability, these tools should be extended beyond the LLaMA family to support a broader range of models, such as the DeepSeek family and the Gemma family, among others.

Regarding \textbf{security risks} in edge LLM deployment, existing research remains limited. Given the heterogeneous and dynamic nature of edge computing environments, potential attack vectors are diverse and complex. One promising approach is executing models within a TEE, which provides hardware-based isolation for secure model inference. However, this method introduces additional challenges, as it requires specialized hardware support and remains vulnerable to various security threats~\cite{cerdeira2020sok}, including side-channel attacks such as CipherFix attacks~\cite{wichelmann2023cipherfix} and cache side-channel attacks~\cite{zili2023cache}. Therefore, future research on \textbf{secure TEE-based LLM deployment} should focus on two key aspects: \textit{(a)} ensuring compatibility across diverse hardware architectures to promote widespread adoption, and \textit{(b)} developing robust defense mechanisms against known vulnerabilities, such as side-channel attacks~\cite{wichelmann2023cipherfix, zili2023cache}, to enhance the security and reliability of on-device LLMs.

\subsection{Hybrid Deployment}

\subsubsection{Research Status}

Hybrid deployment, such as edge-cloud collaborative computing, integrates the advantages of both edge and cloud deployment, offering additional benefits. On one hand, it enhances the flexibility of model deployment and task execution—computationally intensive tasks can be offloaded to cloud clusters, whereas lightweight tasks can be processed on edge devices. On the other hand, it broadens the application scenarios of LLMs, enabling cross-regional task execution and real-time decision-making in latency-sensitive applications.

\textbf{Device collaboration.}  
Effective collaboration between cloud and edge devices requires dynamic task partitioning and computation offloading while balancing inference latency, bandwidth constraints, and data privacy~\cite{zhang2024edgeshard,he2024large,yang2024efficient}. He \ea{}~\cite{he2024large} proposed an active inference approach using reinforcement learning for resource scheduling in cloud-edge LLM inference. CE-CoLLM~\cite{jin2024collm}, a cloud-edge collaborative inference framework, employs early-exit mechanisms, a cloud context manager, and quantization to reduce high communication overhead, achieving both low-latency edge standalone inference and high-accuracy cloud-edge collaborative inference. Additionally, Hao \ea{}~\cite{hao2024hybrid} introduced a hybrid inference method that leverages small models on edge devices in conjunction with large cloud-based models to enhance inference performance.

\textbf{Data security.}  
Hybrid deployment involves extensive data exchange, increasing the risk of data leakage. Common privacy-preserving techniques include federated learning~\cite{chen2023federated,kuang2024federatedscope}, differential privacy~\cite{shi2022just,charles2024fine,novado2024multi}, and split learning~\cite{qu2025mobile}. Federated learning protects user data by enabling local model training while updating a global model in the cloud. Differential privacy introduces noise into data to obscure sensitive information, whereas split learning transmits only partial computation results to the cloud, thereby minimizing data exposure. However, there still are privacy risks, as adversarial servers may attempt to reconstruct users’ original data~\cite{qu2025mobile}. Furthermore, as previously discussed, attacks such as leveraging soft prompt to expose LLM training data could also manifest in hybrid deployment settings. Ensuring data security in hybrid deployments remains an open research challenge that requires further investigation.

\subsubsection{Challenges}

Hybrid deployment introduces additional \textbf{challenges} beyond those encountered in cluster and edge deployment. 
\textbf{Device collaboration.}  
Unlike cluster deployment, hybrid deployment requires efficient coordination between edge devices and cloud servers. The heterogeneity of edge devices, each with varying computational capabilities, further complicates scheduling, making task offloading and resource allocation more challenging. Additionally, optimizing these processes must account for inference latency, bandwidth constraints, and dynamic workload distribution, increasing the complexity of system coordination.
\textbf{Data security.}  
Compared to edge deployment, hybrid deployment involves frequent data exchanges with cloud servers, heightening the risk of data exposure. Unlike in cluster deployment, this risk is made worse by the diverse and less controlled nature of edge environments. Moreover, since both cloud servers and edge devices participate in computations, either party could act maliciously, making security threats more unpredictable and the deployment environment more complex.

\subsubsection{Road Ahead}

To address the challenges associated with \textbf{device collaboration}, future research can explore \textbf{collaborative optimization schemes} from three key perspectives: low-latency communication, performance optimization, and intelligent scheduling. For \textbf{low-latency communication}, advancements in networking and data transmission technologies are crucial for accelerating information exchange within collaborative networks and minimizing communication delays. From a SE perspective, optimizing the inference process itself can significantly enhance communication efficiency between cloud and edge devices. Future solutions could integrate techniques such as vector database caching~\cite{yao2024velo} and MoE architectures~\cite{jin2025moe} to reduce latency and improve inference performance. 
Regarding \textbf{performance optimization}, traditional LLM enhancement techniques (e.g., RAG, MoE, and prompt engineering~\cite{qiao2025prismprompt,jin2025moe}) can be leveraged to improve hybrid inference efficiency. Additionally, novel strategies, such as constraint satisfaction mechanisms for complex decision-making in edge-cloud collaboration~\cite{yang2024perllm} and the integration of quantization with efficient local inference methods~\cite{ray2024llmedge}, may further enhance the computational capabilities of edge devices.  
For \textbf{intelligent scheduling}, future research can focus on developing more adaptive task and resource allocation strategies~\cite{yang2024perllm,zhang2024edgeshard}. These strategies should ensure system robustness by dynamically adjusting to inference failures, resource constraints, and evolving workloads in real-time.  

To enhance \textbf{data security}, \textbf{cryptographic methods} such as HE, zero-knowledge proofs (ZKP)~\cite{sun2024zkllm}, MPC~\cite{zobaed2025confidential}, and blockchain-based security mechanisms~\cite{li2021security,wang2022integrating} present potential solutions. However, these cryptographic methods significantly increase the computational overhead of LLM inference, which limits their applicability on edge devices with limited resources. Alternative approaches such as \textbf{federated learning}~\cite{ye2024openfedllm,kuang2024federatedscope} and confidential computing~\cite{zobaed2025confidential} have gained attention due to their compatibility with hybrid deployment environments, with differential privacy playing a critical role in federated learning.  
Inspired by this, a promising direction involves injecting noise into prompts while ensuring that LLMs can still correctly interpret the intended information. Since prompts primarily consist of natural language, a \textbf{prompt encoding-decoding technique} could be explored to transform prompts into structured representations for noise injection, followed by decoding them back into natural language when needed. A related approach is prompt obfuscation~\cite{pape2024prompt,lin2025emojiprompt}, which can also protect sensitive information within prompts from being extracted or exploited by adversarial entities. These techniques enhance prompt security while preserving the effectiveness of LLM interactions.

\section{Maintenance and Evolution}\label{sec:maintenance_evolution}

\subsection{Research Status}

\begin{figure}[htbp]  
    \centering     
    \includegraphics[width = \textwidth]{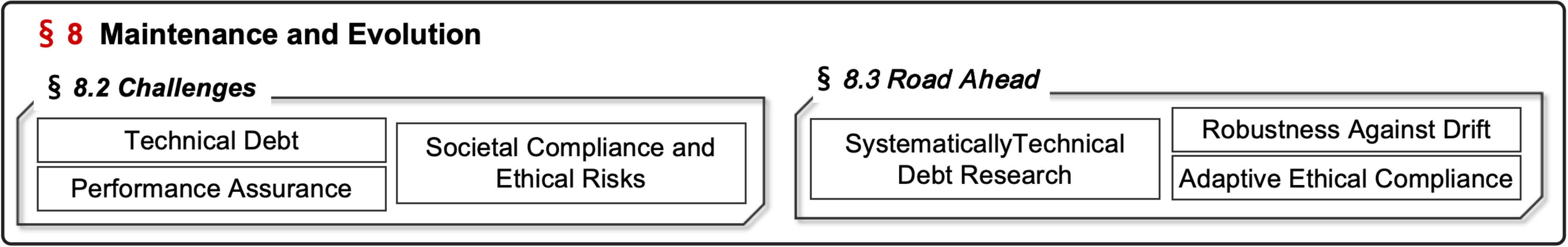}  
    \caption{Challenges and Road Ahead in  \S\ref{sec:maintenance_evolution} Maintenance and Evolution.}
    \label{fig:maintenance_road_ahead}

\end{figure}

LLM maintenance and evolution encompass the ongoing operation, monitoring, and upgrading of models post-deployment, ensuring stable inference, addressing emerging issues, and continuously improving model performance. Unlike the development and enhancement phases, which primarily focus on model training and fine-tuning, maintenance and evolution require long-term management strategies to sustain the efficiency, reliability, and compliance of LLMs in real-world applications. We list the challenges and potential future directions in \autoref{fig:maintenance_road_ahead}.

The rapid advancement of LLMs often leads to the accumulation of \textbf{technical debt}, as ad-hoc solutions (e.g., memory management, model compression, and attention optimization) are implemented to address short-term challenges~\cite{menshawy2024navigating}. However, these solutions may hinder long-term sustainability by increasing system complexity and maintenance overhead. In addition to these architectural challenges, iterative model updates introduce versioning complexities, including API compatibility issues and dependency conflicts, further increasing maintenance efforts. As Ma \ea{}~\cite{ma2024my} highlighted, APIs may be no longer used during model development, making regression testing to be a crucial concern. While tools such as MLflow Model Registry~\cite{mlflow_model_registry} and PEFT~\cite{huggingface_peft} mitigate some of these challenges, critical gaps remain in quantifying technical debt and designing unified lifecycle frameworks that balance incremental learning with catastrophic forgetting.

Ensuring consistent model performance in dynamic environments requires addressing \textbf{model drift}, such as task~\cite{abdelnabi2024you}, data~\cite{jang2024driftwatch}, semantic~\cite{dzeparoska2024intent,roe2024semantic,santos2024adaptive}, concept~\cite{yang2024adapting}, and knowledge~\cite{fastowski2024understanding} drift, which are common in multi-model collaboration. Existing techniques, such as drift detection (e.g., the Kolmogorov-Smirnov test~\cite {gu2024llms}) and model compression, offer partial solutions which may need full model retraining, remaining computationally expensive, and semantic-level drift is often detected with significant delays. Furthermore, The increasing complexity of multi-model collaboration makes these drifts more severe. Maintaining model performance in such settings requires not only early drift detection but also the development of robust adaptation mechanisms to mitigate its effects. However, current solutions remain insufficient, either relying on costly retraining or failing to address higher-level semantic and knowledge drift, which can cause subtle yet significant deviations in model behavior over time.

Beyond technical considerations, LLMs must also comply with evolving legal frameworks (e.g., the EU AI Act~\cite{edwards2021eu}) and ethical standards, necessitating the development of automated compliance mechanisms that integrate regulatory constraints into model behavior. However, automating compliance remains a significant challenge. While post-hoc filters and bias assessment tools help mitigate immediate risks~\cite{10.1145/3650105.3652294,xu2024pride}, they are insufficient for preventing long-term societal harms, such as the reinforcement of biases from incremental data updates or cross-cultural misalignment~\cite{barclay2024investigating,treude2023she,dai2024bias}. 

Thus, maintaining and evolving LLMs need comprehensive solutions that integrate \textbf{technical, operational, and regulatory} considerations. Future research must explore systematic strategies for managing technical debt, improving drift adaptation mechanisms, and developing more proactive compliance frameworks to ensure that LLMs remain reliable, efficient, and aligned with ethical and legal standards.

\subsection{Challenges}

The maintenance and evolution of LLMs present multifaceted challenges that extend beyond traditional SE paradigms. These challenges can be categorized into three interconnected dimensions, encompassing both technical and societal complexities: \textbf{technical debt}, \textbf{performance assurance}, and \textbf{societal compliance and ethical risks}. 

\textbf{Technical debt.} The rapid advancements in model compression, fine-tuning, and continual learning have led to the accumulation of hidden technical debt. For instance, the increasing complexity of model architectures reduces post-training interpretability, while continual learning may exacerbate rather than mitigate biases~\cite{menshawy2024navigating}. These technical debts pose significant risks to model improvement, inference reliability, and security. However, due to the lack of systematic studies in this area, a comprehensive understanding of LLM technical debt remains elusive, making it hard to develop effective mitigation strategies.

\textbf{Performance assurance.} Model drift has emerged as a critical challenge, often resulting in unexpected inference errors, degraded performance, or even the generation of harmful content. Given the dynamic nature of deployment environments and evolving user interactions, mitigating model drift requires robust adaptation mechanisms. Yet, existing methods remain limited in their ability to detect and counteract drift efficiently, particularly at the semantic level, where subtle but significant changes in model behavior can occur over time.

\textbf{Societal compliance and ethical risks.} As LLMs are increasingly deployed in real-world applications, ensuring their outputs align with ethical and social values is imperative. However, current alignment efforts remain insufficient, as adversarial attacks can manipulate even aligned models into producing undesirable outputs. Furthermore, the emergence of \textit{alignment faking}~\cite{greenblatt2024alignment} raises concerns that models may exhibit alignment during evaluation but deviate in other scenarios, casting doubt on the reliability of existing alignment techniques. Addressing ethical alignment in LLMs thus remains an ongoing and pressing research challenge.

\subsection{Road Ahead}

As of the date of writing, research on model maintenance remains limited, particularly in the context of technical debt. However, it addresses a critical issue that deserves greater attention. For the technical debt of LLM, we suggest that the first step in future research should be to do work like \textbf{systematically technical debt research} to assess its impact, and subsequently develop effective mitigation strategies. Additionally, LLMOps~\cite{diaz2024large} has emerged as a promising paradigm for enabling automated lifecycle management and standardized control in LLM development, effectively mitigating common technical debt issues. For instance, LLMOps facilitates real-time model monitoring and continuous feedback mechanisms, allowing for the timely detection and correction of model degradation or knowledge loss, thereby preventing the long-term accumulation of quality debt. By leveraging automation, standardization, and optimization techniques, LLMOps holds significant potential for addressing various technical debt challenges in industrial LLM applications, making it a compelling research direction.

Similarly, \textbf{robustness against drift} necessitates systematic solutions to mitigate the effects of drift in LLMs. Future research may explore the alignment of multi-modal and multi-model representation spaces, prompt refinement, and techniques for preserving semantic integrity in long and sequentially evolving contexts. Additionally, the development of an automated model evolution framework or a performance monitoring system could provide continuous assessment of LLM performance across tasks, both during regular operation and after knowledge updates. By detecting model drift, these systems would allow early action to fix issues, ensuring reliability and stability throughout the LLM development lifecycle, and preventing unexpected performance drops while maintaining accuracy across different deployment scenarios.

Bias and ethical concerns in LLMs must not be overlooked, making \textbf{adaptive ethical compliance} a crucial area of study. While numerous techniques have been proposed to address societal compliance and ethical risks during LLM training, future research should focus on dynamically and precisely adjusting ethical and bias-related concepts during model maintenance, which may involve integrating knowledge unlearning and continual learning techniques while simultaneously addressing the challenge of \textit{alignment faking}~\cite{greenblatt2024alignment}. Furthermore, as regulatory and ethical constraints on LLMs continue to evolve, translating legal frameworks and ethical norms into enforceable model constraints will be essential for ensuring sustained compliance with regulatory changes and societal expectations. Importantly, these adaptations must be achieved without introducing excessive system complexity, thereby preserving the efficiency and scalability of LLM deployment and operation.



\section{Related work}\label{sec:related_work}

With the rapid advancement of LLMs and their success across various applications, research on LLMs has experienced explosive growth in recent years. To systematically summarize existing achievements and outline future directions, a substantial number of survey studies have emerged. Overall, the existing surveys can be categorized into two groups: those focusing on the fundamental aspects of LLMs and those emphasizing the applications of LLMs in different domains.

On one hand, as LLMs include multiple dimensions such as model architecture, training methodologies, and security evaluation, many existing surveys focus on specific aspects of LLM research. Zhao \ea\cite{zhao2023survey} and Naveed \ea\cite{naveed2023comprehensive} primarily concentrate on the development trajectory of LLMs, providing detailed reviews of key technological advancements and major research milestones. Chang \ea\cite{chang2024survey}, Xu \ea\cite{xu2022systematic}, and Guo \ea\cite{guo2023evaluating} focus specifically on evaluation techniques for LLMs. In addition, as concerns over security risks grow with the increasing scale of models, Yao \ea\cite{yao2024survey}, Wang \ea\cite{wang2024unique}, and Das \ea\cite{das2025security} provide systematic analyses of LLM security issues, covering impact assessment, domain-specific vulnerabilities, and overarching security challenges, respectively.

On the other hand, survey studies focusing on the application of LLMs have also been increasing. In particular, the use of LLMs for SE has emerged as a highly active area of research in recent years. Hou \ea\cite{hou2024large} conducted one of the systematic studies in this field, followed by further investigations by Fan \ea\cite{fan2023large} and Liu \textit{et al.}~\cite{liu2024large}. Beyond SE, several surveys have summarized the applications of LLMs across various industries, such as telecommunications~\cite{zhou2024large}, medicine~\cite{thirunavukarasu2023large}, and education~\cite{kasneci2023chatgpt}. Additionally, research has explored LLMs in roles as human judges (LLM-as-a-judge)\cite{li2024generation,gu2024survey,li2024generation}, as well as in SE subfields such as code generation\cite{jiang2024survey,wang2023review} and code repair~\cite{zhang2024systematic}, alongside comprehensive analyses of LLM applications~\cite{hadi2023survey,kaddour2023challenges}. These studies highlight the critical role of LLMs in today's society and underscore their potential for future development, emphasizing the continuing importance of advancing LLM technologies.

However, existing studies have not provided a systematic analysis of the LLM development lifecycle from an SE perspective. To the best of our knowledge, \textbf{we presents the first comprehensive survey in this work} that examines the LLM development lifecycle through the lens of SE. Our study systematically reviews the key SE challenges associated with LLM development and proposes critical directions for future research, offering valuable insights to guide subsequent work in this emerging area.

\section{Conclusion}\label{sec:conclusion}
This paper presents a comprehensive analysis of the challenges associated with LLMs from an SE perspective. By systematically examining each phase of the LLM development lifecycle, we provide an in-depth review of the current research landscape, identify key challenges, and present future research directions. Our findings provide valuable insights to facilitate further advancements in this field, contributing to the development of more efficient, robust, and scalable LLMs.

\bibliographystyle{ACM-Reference-Format}
\bibliography{acmart}

\end{document}